# A QUANTUM APPROACH TO DARK MATTER

Dr A. D. Ernest

*Physics & Electronics*
*School of Biological, Biomedical & Molecular Sciences*
*University of New England*
*Armidale, NSW 2351, Australia*
*Email: aernest@pobox.une.edu.au*

*Abstract:*

*This work develops and explores a quantum-based theory which may enable the nature and origin of cold dark matter (CDM) to be understood without the need to introduce exotic particles. Despite the excellent successes of Lambda Cold Dark Matter (**LCDM**) cosmology on large scales, it continues to encounter several significant difficulties in its present form, particularly at the galactic cluster scale and below. In addition, crucial to **LCDM** theory is the existence of a stable weakly interacting massive particle (WIMP) which is yet to be detected. Using a quantum approach however, it is possible to predict the existence of certain macroscopic quantum structures that are WIMP-like even when occupied by traditional baryonic particles. These structures function as dark matter candidates for **LCDM** theory on large scales where it has been most successful, and retain the potential to yield observationally compliant predictions on galactic cluster and sub-cluster scales.*

*Relatively pure, high angular momentum, eigenstate solutions obtained from Schrödinger's equation in weak gravity form the structural basis of this quantum approach. They are seen to have no classical analogue, and properties radically different to those of traditional localised matter (whose eigenstate spectra contain negligible quantities of such states). Salient features of some of the more tractable solutions include radiative lifetimes, which often far exceed the age of the universe, and energies and 'sizes' consistent with that expected for galactic halos. This facilitates the existence of sparsely populated structures with negligible electromagnetic emission and an inherent inability to undergo further coalescence or gravitational collapse.*

*Viable structure formation scenarios can be constructed based on the seed potential wells of primordial black holes formed at the $e^+/e^-$ phase transition. Composed of a plethora of relevant eigenstates, these structures potentially have suitable internal density*

*distributions and sufficient capacity to accommodate the required amount of dark matter. Additionally, eigenstate particles interact only negligibly with electromagnetic radiation and other baryonic particles. Traditional matter occupying such states will therefore be both invisible and weakly interacting with 'ordinary' particles, including macroscopic galactic objects. Structure formation, probably occurring before big bang nucleosynthesis (BBN), will introduce significant density inhomogeneity. If so, it should be possible to incorporate structures into universal evolutionary scenarios without significantly disturbing expected BBN ratios. The theory therefore provides a mechanism for dark matter to be primarily baryonic, without compromising the results of WMAP or the measurements of elemental BBN ratios.*

## I. INTRODUCTION AND BACKGROUND

Even before the development of General Relativity, scientists realised the possibility that a body could be 'dark' if its gravity was so strong that not even light could escape. Then in the late 1930's observations by Zwycky[1] showed that some kind of non-luminous matter was not only present in galactic clusters, but a necessary major constituent, required just to keep them stable. Since that time evidence for the existence of dark matter has continued to grow. The rotation velocity curves of stars in galaxies, the behaviour of galaxies in clusters and super-clusters, and the lensing effects of galaxies and galactic clusters on photons reaching us from the most distant parts of the universe, all indicate that the universe contains predominant quantities of non-luminous matter whose influence is felt only through its gravitational effects. Yet despite 65 years of study the nature and origin of this dark matter remains unresolved.

There has been no shortage of suggestions for dark matter candidates. The earliest, simplest and most obvious answer was that dark matter consisted of ordinary familiar matter that was dark – for example dead stars, planetary-sized objects that never initiated thermonuclear reactions and diffuse clouds of hydrogen or dust. When it became clear that these baryonic sources were unlikely to fully account for the vast quantity and distribution of dark matter that must be present in the universe (at least about ten times the amount of visible matter), astronomers turned to more esoteric solutions. Suggestions included black holes (either primordial or formed via very early star deaths) and non-baryonic candidates - hot dark matter (HDM) and cold dark matter (CDM).

The reason for the continued mystery surrounding the nature of dark matter however is not an inability to choose the best candidate, but rather the absence of a candidate that appears compatible with the ever growing observations. It would seem therefore that there is some merit in looking for a totally fresh approach to understanding the dark matter problem. Nevertheless, because the experimental results of WMAP and *n*-body simulation predictions of large scale structure favour some sort of CDM as a primary candidate, this would suggest that, in some way at least, CDM is the appropriate path to follow. (See for example Rees[2])

What problems does CDM cosmology face? CDM (or Lambda CDM ($\Lambda$CDM)) works well on large scales. On cluster scales and below however CDM faces several problems which are not yet resolved. These include density profiles which are too cuspy and a predicted overabundance of satellite galaxies and substructure. Attempts to remedy the problems by introducing mechanisms such as self-interacting dark matter have not been able to resolve the issues (D'Onghia and Burkert[3]). The mechanism in CDM cosmology for formation of large halos is through hierarchical merging. This merging requires time $\sim 10$ billion years. Large galaxies have now been found which are fully formed in less than 1 billion years after the big bang (Rocca-Volmerange *et al.*[4]). Also to date there is no discovered particle suitable for CDM theory. What is needed is a theory that predicts WIMP-like particles that behave differently to the models and *n*-body simulations on the scales where CDM or $\Lambda$CDM fails. Is such a theory possible? This chapter presents an alternative approach based on a logical extension of quantum ideas to develop a self consistent theory of the nature and origin of dark matter. It does not replace CDM but provides a basis for understanding CDM (or $\Lambda$CDM) cosmology in a new context. In this context the theory provides 'particles' compatible with large scale structure evolution in $\Lambda$CDM cosmology that retain weakly interacting massive particle properties within the halo structure on the cluster and galactic scales. Such structures may also better predict halo behaviour and density profiles. Most significantly, in the scenario discussed, all this may be done without the need for exotic particles, but nevertheless do not necessarily conflict with the results from WMAP or Big Bang Nucleosynthesis (BBN) ratios.

The next section deals with the justification for considering a gravitational quantum approach to the dark matter problem. Following this, a simplified formal gravitational quantum equation and its solution is presented. This leads in to an extensive discussion of the various properties of the eigenstate solutions that relate to the dark matter problem. The chapter then goes on to present a possible formation scenario for the quantum "eigenstructures" (the arrays of macroscopic gravitational eigenstates) that are proposed to form the basis of dark matter halos. Lastly the chapter concludes with some further discussion, including possible methods of testing the theory and a summary of the salient features.

## II. THEORY

### A. Scope and Justification

The essence of the present approach is very simple: what happens if traditional quantum theory is applied to gravitational potentials? Present theories of quantum gravity are still poorly developed. This work is not an attempt to provide such a theory but rather to investigate the consequences of incorporating gravitational potentials into quantum theory. This should be possible in all but the strongest gravitational regions (for example near black holes close to the Schwartzschild radius $r_s$). It is not necessarily a foregone conclusion that gravity should behave similarly to other force fields. It has however already been experimentally established that gravity may be treated like any other potential in quantum theory. Recent experiments by Nesvizhevsky *et al.*[5] show this by demonstrating quantisation in a wedge-shaped gravitational well formed from the intersection of a 'hard' potential (a horizontal mirror at one location) with the Earth's uniformly increasing gravitational potential above it. Ultra cold neutrons were projected above the horizontal mirror to travel in what would classically have been parabolic paths. Nesvizhevsky and his co-workers found the neutrons were gravitationally quantised and were able to measure the first four energy levels of the neutrons in the Earth's gravitational field as they cascade through the quantised levels of their arrangement. Significantly, the four energy levels were at the peV level, demonstrating one of the reasons why 'gravitational' eigenstates are difficult to observe in everyday life.

In the cosmological context the effects of quantum theory are generally ignored in all but the earliest times when the observable universe was extremely small. This is understandable because, traditionally, quantum effects are considered to manifest themselves only over atomic scales. Astronomical dark matter however relates to huge structures such as galactic halos and the like, having scales of the order of $10^{21}$ m or more. In what way could quantum theory allow the existence of such macroscopic states and how might they be populated? On one hand the Correspondence Principle states that, on the macroscopic scale, quantum and classical approaches should yield identical results. This principle was put forward in order that any new quantum formalism should asymptotically approach what is observed macroscopically. This does not preclude though, the possibility that quantum theory may lead us to become aware of new and novel macroscopic phenomena that have no classical analogue. In the case of dark matter, if quantum theory predicted the existence of astronomically sized, weakly interacting, 'invisible' states it is hardly surprising that they have not been detected earlier. Indeed the evidence suggests that quantum theory does not "break down" as systems become larger and more complex, but rather that macroscopic quantum effects are, for various reasons, harder to observe. It is not surprising therefore that in the last few decades novel quantum effects have been increasingly demonstrated on macroscopic scales.

Indeed there are many examples of macroscopic quantum non-classical phenomena:

(1) Macroscopic Bose-Einstein condensates demonstrate a type of macroscopic "eigenstructure" formed when up to $10^9$ optically trapped atoms "condense" into the same quantum state (the many body, ground state wavefunction of the assembly). These macroscopic structures, which so far have been created up to 0.3 mm in size, have properties profoundly different from traditional matter (see, for example, Ketterle[6], Stamper-Kurn and Ketterle[7], Greiner *et al*[8], Bloch, Hansch and Esslinger[9] and Inouye, S. *et al*[10]).

(2) A set of experiments has been performed that demonstrates a real macroscopic superposition of quantum states analogous to the famous Schrödinger's cat thought experiment using superconducting quantum interference devices (SQUIDs). (Nakamura *et al*[11], Friedman, *et al*[12] and Blatter, *et al*[13]).

(3) Although also explainable from classical electromagnetic theory, the stellar interferometer essentially also demonstrates quantum coherence of optical wavelengths over macroscopic distances (Hanbury Brown and Twiss[14]).

(4) Perhaps the most dramatic and compelling experiments of all however that demonstrate the macroscopic nature of quantum mechanics are those that have followed on from the tests of the conjectures of Einstein, Podolsky and Rosen[15] through Bell's inequality (Bell[16]). The early experiments of Kocher and Commins[17], S J Freedman and J F Clauser[18] and Aspect *et al*[19] have prompted the development of an entire branch of physics involving aspects of quantum connectivity, entanglement and quantum cryptography. In these experiments the measurement of the state of one member of a quantum system consisting of an entangled photon pair defines the state of the other member, even though the individual photons of the system are separated by macroscopic distances of metres or more.(see Tittel *et al*[20])

Even more significant, and of relevance to the present discussion, is that separate observations of individual photons can be carried out sufficiently close to one another in time that any mutual influence that the measurement of one photon has on the other is outside the realm of Einstein locality (Weihs *et al*[21]). This 'superluminal quantum connectivity' has more recently been demonstrated to an extent that would, in non-quantum relativistic physics, be the equivalent of greater than $10^6$ $c$ in the experimental frame and over $10^4$ $c$ in the Cosmic Microwave Background frame (Zbinden *et al*[22] and Scarani *et al*[23]). Further experiments have demonstrated entangled photon state systems stretching up to over a hundred kilometres (Stucki *et al*[24], Kosaka *et al*[25]) and more recently, macroscopically entangled atoms and molecules (Simon and Irvine[26]). The inherent implication of the results of these experiments emphasises the behaviour of entangled states as single macroscopic systems.

In summary, although debate may continue over whether or not quantum wavefunctions themselves represent anything real, there is clear evidence that quantum effects such as coherence and quantum collapse can at least occur over many kilometres and at superluminal speeds. Clearly recent experiments demonstrate that the notion that quantum effects and quantum coherence can act only over atomic scales is a preconceived and possibly mistaken idea. It has grown up with quantum theory rather than a being a deduction that follows from it. Such a notion should therefore not deny the existence of macroscopic quantum eigenstates. The concepts behind some of these quantum phenomena turn out to be valuable in developing the formation and evolutionary scenario of gravitational eigenstates.

**B. A gravitational quantum equation**

It is a simple matter to recast the Schrödinger equation for two particles undergoing weak gravitational interaction into the form

$$-\frac{\hbar^2}{2\boldsymbol{m}}\nabla^2 \boldsymbol{y} - \frac{GmM}{r}\boldsymbol{y} = i\hbar \frac{\partial \boldsymbol{y}}{\partial t} \qquad (1)$$

where $M$ and $m$ are the masses of the two particles, and $\boldsymbol{m}$ is the reduced mass. Solutions to this equation yield the energies and wave functions of the gravitational eigenstates and are immediately obtained by comparison with those for the hydrogen atom (see any quantum text, for example, Schiff[27]) and give the energy eigenvalues $E_n$ as

$$E_n = -\frac{\boldsymbol{m}G^2 m^2 M^2}{2\hbar^2 n^2} \qquad (2)$$

Introducing the parameter $b_0 = \frac{\hbar^2}{G\boldsymbol{m}mM}$, the corresponding eigenfunctions $u_n(\mathbf{r},t)$ are

$$u_{n,l,m}(\mathbf{r},t) = R_{n,l}(r) Y_{l,m}(\theta,\phi) \qquad (3)$$

where $Y_{l,m}(\theta,\phi)$ are the normalised spherical harmonics and

$$R_{n,l}(r) = N_{nl} \left(\frac{2r}{nb_0}\right)^{l} \exp\left(-\frac{r}{nb_0}\right) L_{n-l-1}^{2l+1}\left(\frac{2r}{nb_0}\right). \qquad (4)$$

In equation (4)

$$N_{nl} = \left\{ \left(\frac{2}{nb_0}\right)^3 \frac{(n-l-1)!}{2n\,(n+l)!} \right\}^{\frac{1}{2}}$$

is a normalising constant and

$$L_{n-l-1}^{2l+1}\left(\frac{2r}{nb_0}\right) = (n+l)! \sum_{k=0}^{n-l-1} \frac{(-1)^{k+2l}\left(\frac{2r}{nb_0}\right)^k}{(n-l-1-k)!(2l+1+k)!k!} \qquad (5)$$

are the generalised Laguerre polynomials in their standard form.

### III. MACROSCOPIC GRAVITATIONAL EIGENSTATES

**A. Eigenstate properties**

Whilst the mathematical formulation of gravitational eigenstates is initially very straightforward, the explanation of dark matter in terms of a significant number of filled eigenstates on the macroscopic or astronomical scale is conditional on several factors:
- The existence of suitable gravitational potentials of sufficient depth and range
- Eigenstate radii up to the size of the galactic halo
- An inability to undergo collapse
- State eigenvalue energies, spacings and wavefunctions consistent with quantum rules for stability
- A sufficient number of suitable states to accommodate the halo mass
- Properties in keeping with those of dark matter, that is, extremely long radiative decay times and weakly interacting with photons and traditional matter
- Generation of density profiles consistent with overall observed rotation curves
- Consistency with other current astrophysical observations
- A suitable formation mechanism

Most attempts to produce a gravity-bound quantum-like system yield nonsense in terms of satisfying the criteria above. Using equations (2) - (4) and ignoring any attraction other than gravitation, two neutrons for example would have their lowest eigenstate spread over a distance greater than that of the galaxy and the energy eigenvalue of the system would be about $-10^{-69}$ eV, that is, essentially unbound. On the other hand, two 1 kg "particles" would have their lowest energy state at $-10^{65}$ eV but a wave function of extent of only $10^{-58}$ m. It is possible however to envisage several situations where, by choosing the appropriate mass, macroscopic gravitational wave functions can be formed with sensible energies. This requires large principal quantum numbers. For example, two $10^{-2}$ kg "particles" with $n = 10^{26}$ and $l = 10^{26} - 1$ would have a binding energy of about 6 keV and an effective range of around 3 m.

The traditional macroscopic situation of two $10^{-2}$ kg masses orbiting at 3 m however is clearly not such an eigenstate: there are no eigenfunctions having this degree of localisation in the $f$ coordinate and the probability density is definitely not stationary. A position measurement on an eigenstate would not yield a value that is temporally predictable as in the case of the orbiting particles. This example highlights the essence of the macroscopic eigenstructure concept: that a gravitational macroscopic stationary state represents a fundamentally different kind of structure from the traditional (extremely mixed) localised orbiting particle structure. In quantum mechanical terms, an orbiting particle is an extremely localised wave packet. In the quantum representation of macroscopically localised particles, the wave packets representing them are so mixed that they contain only a negligible component of any one single pure state.

Observing or even producing such a gravitational macroscopic eigenstate with two such masses in the laboratory would be virtually impossible. The probability of two orbiting masses making a chance transition to this or similar states from two traditional orbiting masses is prohibitively small, because the size of the relevant overlap integral is negligibly small. There are two significant reasons for this. One is that a single eigenstate necessarily represents an infinitesimally small fraction of the total eigenspectrum making up any classically orbiting particle arrangement. The other relates to the highly symmetric oscillatory properties of the large $n$ Laguerre polynomials and large $l$ spherical harmonic functions to be discussed at greater length later in this chapter.

**B. Eigenstate particles**

Theoretically any mass can be put in an eigenstate. Gravitational eigenstructures could conceivably consist of any stable particles such as protons, electrons, atoms or something more exotic (for example weakly interacting particles, such as axions or neutralinos). As will be seen in what follows, however, the eigenstate theory does not require particles to be intrinsically weakly interacting: the eigenstates themselves being considered here naturally possess weakly interacting properties. Thus the direct predicted existence by quantum theory of bound gravitational states that are weakly interacting opens the door to the possibility that exotic particles and exotic physics are not needed to explain dark matter. Traditional particles and quantum physics is enough.

In the formation scenario to be presented later in this chapter, dark matter halos formed in compact structures from baryons and electrons much earlier than in the standard CDM models. The eigenstate particles are thought to consist of protons, electrons and possibly neutrons and atoms, depending on the interaction properties of these particles in eigenstates. Because formation is thought to occur early, it is envisaged that eigenstates would initially consist of protons and electrons. It turns out that because of the quantum numbers involved, it seems likely that eigenstates might remain as separate proton and electron states. This relates to arguments about the value of the relevant overlap integrals. Even so, the possibility of proton-electron recombination at earlier times does need further consideration. (Consideration of the classical electron-proton recombination rate at present day low halo particle densities suggests that eigenstates would not recombine significantly at the present day.)

When considering electrons and protons coexisting in eigenstates at present day halo densities a simple classical argument shows that it is not necessary to introduce electrical terms into equation (1). The electrical potential energy of a single charge $q$ embedded in a uniform array of positive and negative charges $q_j$ maybe estimated by considering the potential energy of a regular array of alternating positive and negative charges analogous to ionic arrays in the solid state. The potential energy is of order

$$\frac{1}{4\pi\varepsilon_0} \sum_{\text{allcharges } q_j} \frac{q\, q_j}{r_j} = \frac{\alpha\, q^2}{4\pi\varepsilon_0\, s} \qquad (6)$$

where $\alpha$ is the Madelung constant (see any text on solid state physics such as Kittel[28]) and $s$ is an average separation distance analogous to the lattice constant. At an average particle eigenstate density of $\sim 10^6$ m$^{-3}$ equation (6) gives the electrostatic potential energy of a typical proton as $\sim 10^{10}$ times smaller than its gravitational potential energy.

Lastly, it is sensible to ask why the properties of a group of particles in eigenstates should be any different from a group of free particles. This is an important question because, since the mathematics predicts otherwise, it should be possible to see how this might physically come about. After all, free particles may themselves be written as sums of complete sets of eigenstate functions and one would expect their properties to be averages of their eigenstate spectra. This is indeed the case but it is precisely because of this, and because any of the pure eigenstates being considered represents such a small fraction of any 'localised particle' quantum state, that one cannot draw conclusions about properties of the individual eigenstates from the properties of localised particles. One should not expect a gravitational eigenstate to have similar properties to those of a free particle any more than an electron in a hydrogenic eigenstate would exhibit similar properties to a free electron, even despite the fact that, in dealing with macroscopic quantum systems one is 'further' away from the uncertainty limit ($\Delta x \, \Delta p \geq \hbar/2$).

## C. Halo eigenstates – radius and energy considerations

An approximation is made (analogous to the electrostatic approximation in the Hartree theory of atoms) by treating each particle within the halo as acting independently and subject to a single central gravitational potential produced by the mass contained within its spherical radius. This 'two particle' approximation reduces mathematical complexity but retains the important features of the individual eigenstates. It is also assumed that the many dynamic and often violent galactic processes have a negligible effect. The justification is based on analogy with the equivalent adiabatic one used with electronic and vibrational motions in atomic physics. The particle has a relatively small mass and the rotational period of the wavefunction in complex space (related to $\exp(-iEt/\hbar)$) is short compared to macroscopic galactic processes. In principle an eigenstate wavefunction will adjust to the relatively "slow" galactic changes similar to electronic wavefunctions adjusting to atomic collisions and vibrations of their ionic cores. This should also mean that, like their electronic counterparts, particles in pure gravitational eigenstates should remain relatively pure. Of course it could be argued that the types of macroscopic wavefunction self-adjustments described here have potential to violate relativity and result in causality issues, but this is still an open question and the reader is referred to the superluminal quantum transformations already experimentally observed in the Bell-type experiments of Zbinden *et al*[22] and Scarani *et al*[23] discussed earlier.

A particle at or near the 'edge' of the halo, ($r \sim 6.5 \times 10^{21}$ m, Alcock[29]) will see almost the total halo mass *M* of say $2 \times 10^{42}$ kg (Alcock suggests values up to $6 \times 10^{42}$ kg). It is immediately obvious from considerations of the large classical angular momentum involved that large quantum numbers ($n, l > 10^{30}$) will be involved in the eigenstate functions. Even if only the very high *n* values are occupied, these large numbers provide plenty of pure states to accommodate the mass of the galactic halo with traditional matter. At such high *n* values, the Laguerre polynomial and spherical harmonic expansions will, except for a few specific cases, involve so many terms that exact expressions are impossible. It is possible however to develop suitable approximations for various useful properties of these functions, such as the probability distribution $\psi(r,\theta,\phi)^*\psi(r,\theta,\phi)$. The dependencies on $\theta$ and $\phi$ are ignored for the moment. Consideration of the radial component $(R_{n,l}\, r)^2$ of $\psi(r,\theta,\phi)^*\psi(r,\theta,\phi)\,dV$ is useful because it gives the effective range and spatial oscillation frequency of an individual eigenfunction in the radial direction, and ultimately leads to radial dark matter density profiles.

It turns out to be useful to introduce a quantum number $p$ such that $p \equiv n - l$. When the function $(R_{n,l}\, r)$ has an angular momentum quantum number $l = n - p$ the function $(R_{n,l}\, r)$ has $p$ turning points and the series representing $R_{n,l}\, r$ contains $p$ terms. If $p$ is small (that is $l$ is close to $n$) it is therefore possible to write down $(R_{n,l}\, r)^2$ exactly. For example, when $n \gg 1$ and $p = 1$ (i.e. $l = n-1$), $(R_{n,l}\, r)^2$ is a narrow single peaked function of the form

$$(R_{n,n-1}\, r)^2 = \frac{4^n \exp\left(-\frac{2r}{n b_0}\right)\left(\frac{2r}{n b_0}\right)^{2n}}{n^2 b_0\, r(2n-1)!} \quad (7)$$

with the peak position $r_{peak} = b_0\left(n^2 - n/2\right)$ and width $\Delta\left[(R_{n,n-1}r)^2\right]_{cw} = b_0 n \sqrt{(2n-1)}$ (see Appendix A). As $p$ increases, three things happen to the eigenfunction. Firstly, the radial extent of $(R_{n,l}\, r)^2$ increases. Secondly, it contains more oscillations and thirdly, it extends beyond the vicinity of $r_{peak} = b_0\left(n^2 - n/2\right)$. By considering truncated series approximations to $(R_{n,l}\, r)^2$ it is possible to estimate the maximum range of $(R_{n,l}\, r)^2$ as $p$ approaches $n$ (that is $l = 0$). For $p = n$, $(R_{n,l}\, r)^2$ has a significant magnitude even down to $r \sim 0$ and extends out to $r \approx 2 b_0 n^2$ (
Appendix A).

The extent of $(R_{n,l}\, r)^2$ may also be estimated from the position of its zeros. It is possible to obtain approximations to these zeros for arbitrary values of $n$ and $p$ without it actually being necessary to know the explicit polynomial form of $(R_{n,l}\, r)^2$. This procedure is outlined in Appendix B. From the approximate positions of the first and last zeros the range of $(R_{n,l}\, r)^2$ may be obtained. From the differences in position of two adjacent zeros the spatial oscillation frequency of $(R_{n,l}\, r)^2$ at any position may also be obtained. This will turn out to be of importance in estimating the value of overlap integrals and other properties of the eigenstates.

The physical meaning of these results may be summarised as follows. When the principal quantum number $n$ is large, the radial eigenfunctions have an extremely sharply defined radial width that depends on $p$. Outside the range of this width the amplitude of the eigenfunction is essentially zero. When $p$ is low (high angular momentum) the width is small. As $p$ increases the width increases but the spatial oscillation frequency increases at a faster rate so that higher $p$ valued eigenfunctions have much more closely spaced oscillations than lower $p$ valued eigenfunctions. As in the theory of atoms the probability density is completely unlocalised in the $f$ direction for all eigenfunctions. For any value of $p$ the sum of eigenfunctions over all $m$ states also gives a set of states that is completely unlocalised in the $q$ direction. Hence as in the atomic case, for given values of $n$ and $p$, the multiplicity of states over all $m$ values is always spherically symmetric.

Whether viewed from classical or quantum physics, gravity must ultimately bind a dark matter particle to the halo. What is the minimum binding energy that a dark matter particle must possess in order to keep it stably bound to the galactic halo? At what distance will a particle have this minimum binding energy? Theoretically, for an isolated halo in an empty universe there is no limit, just as is the case for an electron in an eigenstate around an isolated hydrogen nucleus. In reality of course the presence of significant background energy (for example, the CMB) and the presence of other nearby halo wells limit the size and mass of any individual halo. This is important because knowledge of the highest energy level, along with the halo mass $M$ and its maximum radius $r_{max}$ can help establish a minimum mass for the eigenstate particle. This can be seen from equation (2) and the relation $r \sim k n^2 b_0$ where $k$ is ~ 1 or 2, which combine to give $E_n \sim GmM/(2kr_n)$ or $m \sim 2k r_{max} E_{max}/(GM)$ where $E_{max}$ is the energy of the highest eigenstate and $r_{max}$ its corresponding radius. Background energy comes mainly from the abundant source of photons in the cosmic microwave background radiation (CMB), ~ $3 \times 10^{20}$ photons per cubic meter with a present day energy spectrum peaked at ~ $1 \times 10^{-3}$ eV per photon. This might seem to be an appropriate energy to use as a bound for the highest eigenstate energy level, but of course this also depends on the rate of interaction with the states (that is whether equilibrium has been established or not). This will depend on the density of photons and on the relevant overlap integrals. As already mentioned, in the scenario being presented here, dark matter halos were principally formed substantially before decoupling ($z \sim 10^3$), when photon energies and number densities were much greater. As the universe ages the background energy and photon number density drops (which favours eigenstate formation) but the particle density also drops which reduces the probability of eigenstate formation. At some point there must be a trade off between falling photon energies and densities and falling particle densities. Given the galactic mass of $2 \times 10^{42}$ kg and halo radius of $6.5 \times 10^{21}$ m, a minimum eigenstate binding energy of 0.001 eV implies a minimum eigenstate particle mass of ~ $4 \times 10^{-32}$ kg, while a minimum eigenstate binding energy of 1 eV implies a minimum particle mass of $3 \times 10^{-29}$ kg. A complementary way of looking at this is to consider the radius of the highest energy wavefunction that could reasonably exist. Taking the minimum binding energy as between 0.001 eV and 1 eV gives the maximum radius for proton wavefunction between $\sim 7\times 10^{26}$ and $\sim 7\times 10^{23}$ m, and between $4\times 10^{23}$ and $\sim 4\times 10^{20}$ m for an electron wavefunction. For masses $\sim 1/1000$ of the electron mass the radius is between $\sim 3\times 10^{20}$ and $\sim 4\times 10^{17}$ m, i.e. much smaller than the expected halo radius. Similar arguments can be made using classical considerations. Such estimates ultimately rule out the very light particles such as axions from being dominant eigenstate particles, since this argument shows that particles lighter than $\sim 1/1000$ of the electron mass cannot produce a large enough halo and remain bound. Likewise if the dark matter particles were much heavier than protons, one might have expected halos to be of greater radial extent than those observed in galaxies and galactic clusters. In other words the mass of the eigenstate particle helps determine why clusters and galaxies are the size they are. Traditional baryonic matter fits well into these order of magnitude mass estimates with the radius of the estimated highest energy level coincidently agreeing with the expected maximum halo radius.

**D. Radiative decay and halo stability**

Essential to dark matter is that it is dark (that is, it must not emit radiation). Fundamental to gravitational eigenstates as dark matter candidates is therefore that they should possess a long radiative lifetime. Long radiative lifetimes also imply potential for eigenstate stability and weak interaction rates with photons (i.e. transparency) and particles, through the involvement of the relevant overlap integrals. In traditional atomic physics the high lying, high angular momentum states (high *n*, *l*) tend to have the longest lifetimes so that, if stable, long lived gravitational eigenstates exist, they are likely to be those with correspondingly high *n*, *l*. Fig. 1 diagrammatically shows the high *n*, high *l* states being considered here. We again use the parameter $p = n - \ell$. Each diagonal line represents a set of states that all have a common value of $p$, starting from the left with $p = 1$. Although there are almost certainly other states that have long lifetimes, the thrust of the present work is to demonstrate that at least this group of states and chiefly this group of states satisfies the requirements for dark matter candidature. Since for *any* fixed values of *l* and *n* the sum over all possible *z*-projection sub-states *m* yields a spherically symmetric probability density, each dot on Fig. 1 represents a set of states with combined spherical symmetry, provided all *m* are included.

Radiative decay of eigenstates could take place via the emission of gravitational and/or electromagnetic wave radiation if the eigenstate particles are charged. There is no full quantum theory of gravity as yet. Nevertheless it can be shown using a classical analogy that, in the case of charged particles, radiative decay through gravitational wave radiation is almost certainly insignificant compared to that produced through electromagnetic radiation. From classical electromagnetic theory, equations for the electric field strength $\mathbf{E_{rad}}$ of the electromagnetic radiation field, and the total electromagnetic radiated power $P$, found from integrating the Poynting vector $\mathbf{S}$ over a spherical surface of radius $r$, are given respectively by, in the low velocity case (any classical electromagnetism text such as Panofsky and Phillips[30])

$$\mathbf{E_{rad}} = \frac{e}{4\pi\varepsilon_0 r^3 c^2} \mathbf{r} \times (\mathbf{r} \times \dot{\mathbf{u}}) \tag{8}$$

and

$$P = \frac{e^2 \dot{u}^2}{6\pi\varepsilon_0 c^3}. \tag{9}$$

Here, $\dot{\mathbf{u}}$ is the acceleration of the particle, r is the radius of the spherical surface and the other terms have their usual meanings. Using a Newtonian approach to avoid introducing concepts from general relativity, the equivalent expressions for a gravitational wave radiation field $\mathbf{G_{rad}}$ and a gravitational radiated power $P$, may be found from integrating the gravitational equivalent of the Poynting vector over a spherical surface radius $r$, in an analogous way, that is

$$\mathbf{G_{rad}} = \frac{Gm}{r^3 c^2} \mathbf{r} \times (\mathbf{r} \times \dot{\mathbf{u}}) \tag{10}$$

and

$$P = \frac{2Gm^2 \dot{u}^2}{3c^3}. \tag{11}$$

Comparing the value of the term $e^2/6\pi\varepsilon_0$ of equation (9) with $2Gm^2/3$ in equation (11) for a proton shows that the rate of energy loss via electromagnetic radiation is about $10^{37}$ times larger than that via gravitational radiation.

The foregoing calculation suggests that it is only necessary here to consider transitions occurring via spontaneous eigenstate decay of charged rather than neutral particles and then only those transitions producing electromagnetic radiation rather than gravitational radiation. This allows equations analogous to those for atomic physics to be used provided that certain similar assumptions are made about the distribution of the radiation field (see for example discussions on spontaneous decay in Corney[31]). The transition probability $A_{i,f}$ ($\equiv$ Einstein $A$ coefficient) for a given state transition $n_i \to n_f$ is given by (Corney[31])

$$A_{i,f} = \frac{w_{if}^3 |\langle f|\mathbf{r}|i\rangle|^2}{3e_0 p \hbar c^3} = \frac{w_{if}^3 p_{if}^2}{3e_0 p \hbar c^3} \tag{12}$$

where $w_{if} = \left(mG^2 m_p^2 M^2 / (2\hbar^2)\right)\left(1/n_f^2 - 1/n_i^2\right)$ is the angular frequency corresponding to the transition $i$ to $f$, $m$ is the reduced mass, $p_{if}$ (not to be confused with the quantum parameter $p = n - l$ introduced earlier) is the corresponding dipole matrix element for the transition $i$ to $f$, and the other symbols have their normal meanings. When a state $i$ has several decay channels, the reciprocal state lifetime $1/t_i$ is then the sum of the transition probabilities over all possible decay channels, $k$: $1/t_i = A_i = \sum_k A_{i,k}$. The major difficulty in calculating $1/t_i$ is finding the values of $p_{if}$ in equation (12).

As explained in
Appendix A there are usually so many terms in the form of the eigenfunctions that it is usually not possible to write down full expressions for quantities such as the dipole matrix elements. An outline of the procedure for the calculation of some of the relevant $p_{if}$ matrix elements that are mathematically more tractable is given in Appendix C. Explicitly $p_{if}$ is written as:

$$p_{if} = \left| \int_0^\infty \int_0^p \int_0^{2p} R_{n_f l_f}^* Y_{l_f m_f}^* e\mathbf{r} R_{n_i l_i} Y_{l_i m_i} r^2 \sin(q)\cos(f)\sin(q)\, df\, dq\, dr \right| \tag{13}$$
$$= \sqrt{\left(p_{ifx}^2 + p_{ify}^2 + p_{ifz}^2\right)}$$

where $p_{ifx}$, $p_{ify}$ and $p_{ifz}$ are the $x$, $y$ and $z$ components of the vector integral inside the modulus sign in equation (13). Dipole radiative decay occurs via transitions involving $\Delta m = 0$ (implying $p_{ifx} = p_{ify} = 0$) or $\Delta m = \pm 1$ (implying $p_{ifz} = 0$) and $\Delta l = \pm 1$ (implying transitions must take place between adjacent columns in Fig. 1). It is possible to calculate upper limits to the size of the angular components of $p_{ifx}$, $p_{ify}$ and $p_{ifz}$. In general the size of all of the angular integrals is small (of the order of unity), and for many specialised cases can be shown to converge to zero for large $l$ (Appendix C).

The radial integral component is a critical factor in determining a transition rate. Appendix C outlines the derivation of a general expression for calculation of the radial integral (with the dipole charge term included). For some types of transitions like those taking place along any diagonal such as $A \to A'$, $B \to B'$, $C \to C'$ etc. in Fig. 1, the approximate value of this expression is easily calculated as $\approx eb_0 n^2$ whenever $n \gg p \, (= n-l)$.

Transitions that move from one diagonal to another like $B \to B''$ involve a "shape change" in the Laguerre polynomial and the radial overlap integral behaves in a way that increases more slowly with $n$. (The radial integral component in this case $\approx eb_0/\sqrt{2} \, n^{\frac{3}{2}}$ for $B \to B''$ type transitions whenever $n \gg p$.)

For the states lying along the left hand diagonal there is only 1 dipole allowed transition: following down the diagonal such that $(\Delta n, \Delta l) = (-1, -1)$. Because there is only one transition, the lifetime of each of these states is simply the inverse transition rate. As an example, consider the transitions of the type A to A′ in Fig. 1. These have $n_f = n_i - 1$ and $p = 1$. Taking a central galactic mass of $2 \times 10^{42}$ kg, and a value for $n_i$ of $1.73 \times 10^{34}$, a proton wave function would have a radius of about $6.0 \times 10^{21}$ m, a radial spread of around $6.4 \times 10^4$ m and a radiative lifetime of $1.6 \times 10^{25}$ s. This lifetime is extremely long (the age of the universe is $\sim 5 \times 10^{17}$ s) and the state is effectively frozen in time. To decay to $n = 1.0 \times 10^{34}$ requires shuffling down through $\sim 7 \times 10^{33}$ transitions or an approximate total time of $\sim 10^{59}$ s ($\sim 3 \times 10^{51}$ years). For electrons, the principal quantum number that gives a radius of $6.0 \times 10^{21}$ m is $n_i \approx 7.7 \times 10^{30}$. The wave function spread is about $3.0 \times 10^6$ m and the single transition lifetime is about $3.0 \times 10^{25}$ s.

$p_{if}$ varies enormously for different individual transitions. For example, consider the result $eb_0 n \left(2/(\boldsymbol{p} \, p)\right)^{\frac{1}{4}} (e/2)^{\frac{1}{2}} (p/n)^{\frac{p-3}{2}}$ (derived in Appendix C) for the value of the radial integral component of $p_{if}$ for transitions such as $E \to E'$ of Fig. 1, which originate from a 'deep' state E and end on the $p = 1$ diagonal. TABLE I demonstrates the rapidity with which $p_{if}$ for this particular radial integral can change as the value of $\Delta n = n_i - n_f$ increases. Despite the cubic dependence on $w_{if}$, the resulting transition rate decreases dramatically with increasing $\Delta n$. It is clearly important to be able to calculate (or at least estimate) the values of $p_{if}$ in the general case because of its large variability.

There are two other points to note here. The first is that the transition linewidths are small compared to the energy level spacing. The natural linewidth of a typical high level state $(n \sim 10^{34})$ is $\sim \hbar/t \approx 10^{-40}$ eV, ($t$ = state lifetime) while the energy separation $\Delta E$ between states is $\boldsymbol{m} G^2 m^2 M^2 \Delta n / (\hbar^2 n^3) \sim 4 \times 10^{-32}$ eV. Using the review of Bland-Hawthorn and Freeman[32] and the dark matter density profile described by Munyaneza and Biermann[33] to obtain an overall radial mass function, it is significant to note that the transition rates between individual high $l$ states increase as the eigenstate radius decreases.

In addition there is more opportunity for overlap of the states with higher $p$ values. This reduced stability implies that the density profiles should become significantly shallower, perhaps even flat or hollow, at these lower radii. This means (see next section) that the intrinsic reaction rates with other photons and particles are also increased for the lower $n$, smaller radius states. Furthermore, near the bulge of the galaxy, also photon and particle densities are higher and these factors contribute to a reduction of the gradient of the density profile of eigenstate material in the inner parts of the halo. These and other qualitative arguments suggest that radial density profiles will not exhibit a cusp-like central profile but they have yet to be investigated in a quantitative way. Further analysis is needed to investigate this.

For proton populated states lying on the diagonals $p = 2,3,4...$ etc. there are progressively more decay channels available. States on the $p = 2 \times 10^7$ diagonal will have available $\sim 4 \times 10^7$ decay channels. If similar transition rates are assumed on average for all channels, then the lifetimes of such states would be $4 \times 10^{17}$ s, which is ~ the age of the universe. From this simplistic assumption, one might estimate the available number of sufficiently long lived states to be (summing over a diagonal strip of width $10^7$ in Fig. 1) $\sim 2n^2 \times 10^7 \approx 5 \times 10^{75}$ for protons and $\approx 9 \times 10^{69}$ for electrons. Assuming that the lower figure for electrons limits the halo size and assuming charge neutrality, this gives at least a mass capacity of these long lived, high $(n,l)$ states as $\sim 6 \times 10^{42}$ kg. This is a crude estimate of the expected capacity and there needs to be some explanation of why the inner states are not populated or have not collapsed into the central regions of the galaxy. The formation and evolutionary scenario described later in this chapter discusses the effects of stimulated transitions on the eigenstates and it may be that these induced transitions in the early radiation dominated universe may be capable of expanding the halo and channelling particles into the longest lived states.

It is worth noting the physical reasons behind why the macroscopic halo eigenstates have such long lifetimes compared to their atomic counterparts. When the initial and final $n$ values are close, the frequency is extremely small because the energy levels in the halo eigenstructure are so much closer than in atomic systems. The factor $w_{if}$ in equation (12) is so minute that the decay rate is negligible. When the level difference $\Delta n$ is large however the values of the radial components of the $p_{if}$ integrals can become minutely small. There are several reasons for this. The $R_{n,l} r$ functions have limited radial extents that depend on $n$ and $p(=n-l)$, particularly so, the closer they are to the $p = 1$ diagonal (see Appendix B). Outside the extent of any radial wavefunction its amplitude drops very rapidly to zero. If the initial and final states have wavefunctions that do not have a common overlap region then the $p_{if}$ value will be negligible. Sometimes the $R_{n,l} r$ functions overlap but the extent of one is a small fraction of the other. Sometimes also, as discussed in the final example of Appendix C, the spatial radial oscillation density of one state far exceeds the other. (Since $\Delta n \gg 1$, the requirement that $\Delta l = \pm 1$ implies that $\Delta p \gg 1$, which implies the difference in the number of oscillations of the initial and final states is large.) This results in extremely effective cancellation because the high $n$ Laguerre polynomials are so symmetrical over most of their extent. For example using equation (2) an energy change sufficient to produce a radio-frequency photon of 10 MHz requires $\Delta n > 2\times 10^{21}$. Consider a transition from $n = 1.0\times 10^{34}, p = 2.0\times 10^{21}$ to $n = 1.0\times 10^{34} - 2\times 10^{21} + 1, p = 1$. Calculations using Appendix B show that not only is the radial extent of the initial upper level $\sim 10^{11}$ times greater than the lower level but also that even within the envelope of the lower level, the upper level oscillates symmetrically almost ten billion times.

Lastly it should be noted that *any* dipole radiating, $\Delta n = 2\times 10^{21}, \Delta l = \pm 1$ level change automatically implies that $\Delta p = 2\times 10^{21} \pm 1$. This implies there will be a difference in oscillation number of $\sim 2\times 10^{21}$ between the two states. Even if the spatial oscillation periods of the two states is not great (as may be the case when $l \ll n$) small differences between the radial oscillation periods of the two functions still cause a beating effect - but the long train of successive 0-phase/$p$-phase beats within the integral effectively cancels over the whole integration range. Within the deeper $p$ regions ($l \ll n$) of the eigenstate structure diagram in Fig. 1 there may be resonances that enable decays to proceed, but such decay may have to take place via very high order multi-pole decay and/or at lower $n$ values. The situation is very complex and requires further investigation. It should be noted lastly that any radiative decay that does proceed from the deeper $p$ states of Fig. 1 favours channelling to toward lower $p$, higher $l$ relative to $n$ values. (For example, the deep state D of Fig. 1 has only one decay channel that keeps it on the same $p$ diagonal, all other channels moving it relatively closer to the $p = 1$ diagonal.) This promotes a net migration of states towards the left hand diagonals, that is, to those states that probably possess the longest lifetimes.

Some more mention should be made of the possibility of higher order, multipole decay. In atomic transitions such decays are largely ignored because the corresponding decay rates are much lower than the dipole rates. Indeed the decay rate for an allowed electric quadrupole transition is smaller than that for an allowed dipole transition by a factor which is of the order of $(r/\mathit{l})^2$ where $\mathit{l}$ is the wavelength of the emitted radiation. This factor is about the same for the atomic states as for those in the macroscopic states considered here. Additionally there is the consideration (to be dealt with further in the next section) that the $p_{if}$ value rapidly approaches zero when there is no state overlap. These remarks imply that multipole decay rates ought to be less than the corresponding dipole rate but a detailed study of the implications of multipole radiation needs to carried out and is beyond the scope of the present introductory investigation.

**E. Photon Interactions**

**(a)** *Stimulated Emission and Absorption*

It was shown in the previous section that some of the halo gravitational eigenstates have extremely long radiative decay times. This section discusses the absorption and stimulated emission of radiation. The universe is filled with cosmic microwave background radiation (CMB), and within the vicinity of galaxies there is an additional energy density component of higher frequency radiation from stars and hot gas. Eigenstates can be shifted up or down via absorption or stimulated emission of radiation. An eigenstate could be successively promoted upwards by radiation until it is ultimately gravitationally freed from the galaxy, in a process analogous to the electrical ionisation of an atom.

The probability of absorption of radiation (or stimulated emission) $P_{i,f}$ from a state $i$ to a state $f$ is given by

$$P_{i,f} = \frac{p e^2}{3 e_0 \hbar^2} \left| \langle f | \mathbf{r} | i \rangle \right|^2 r(w_{if}) \tag{14}$$

where $e \left| \langle f | \mathbf{r} | i \rangle \right|$ is the same as the dipole matrix element $p_{if}$ used in equation (12) and $r(w)$ is the spectral energy density per unit angular frequency (see for example, Corney[31]). The Einstein $B$ coefficient $B_{i,k}$ is defined here through the relation $B_{i,f} = r(w_{if})/P_{i,f}$. When $i$ and $f$ are degenerate levels (of degeneracy $g_i$ and $g_f$ respectively) then $B$ must be summed over the final states $m_k$ and averaged over the initial states $m_i$ to give

$$B_{i,f} = \frac{p e^2}{3 e_0 \hbar^2 g_i} \sum_{m_i, m_k} \left| \langle m_k | \mathbf{r} | m_i \rangle \right|^2 = \frac{c^3 p^2}{\hbar w_{if}^3} \frac{g_f}{g_i} A_{f,i} \tag{15}$$

As a first case, consider a $\Delta n = \pm 1$ transition along the $p = 1$ diagonal. The angular frequency $w_{if}$ required to match a $\Delta n = \pm 1$ proton wavefunction shift is $\sim 7 \times 10^{-17}$ rads$^{-1}$ for $n \sim 1.0 \times 10^{34}$. Assuming that $g_i \approx g_f$ since the transitions are from high $l$ to similarly high $l$ levels ($\Delta l = \pm 1$, $\Delta m = 0, \pm 1$) and taking $A_{f,i} \sim 1 \times 10^{-25}$ for the $n$ to $n \pm 1$ transition at $n \sim 10^{34}$ and $r(w_{if}) \sim 7 \times 10^{-82}$ J m$^{-3}$ srad$^{-1}$ for the present day CMB (2.73 K) at this angular frequency, gives the absorption/stimulated emission probability as $B_{i,f} \, r(w_{if}) = \dfrac{p^2 c^3}{\hbar w_{if}^3} \times A_{f,i} \times r(w_{if}) \sim 5 \times 10^2 \text{ s}^{-1}$. If this rate was the same throughout the history of the universe, the maximum change in principal quantum number $n$ over the lifetime of the universe would be $2.5 \times 10^{20}$. This is a negligible change for a state with $n \sim 1.0 \times 10^{34}$.

As a second case, consider the peak value of the present day CMB $r(w)$ of approximately $2.6 \times 10^{-25} \text{ J m}^{-3} \text{ srad}^{-1}$ at an angular frequency of $1.0 \times 10^{12} \text{ s}^{-1}$. Again using $B_{i,f} = \dfrac{c^3 p^2}{\hbar w_{if}^3} \dfrac{g_f}{g_i} A_{f,i}$, with $g_i \approx g_f$, the stimulated absorption/emission probability $B_{i,f} \, r(w_{if})$ at this frequency becomes $\dfrac{c^3 p^2}{\hbar w_{if}^3} \times A_{f,i} \times r(w_{if}) = 0.06 \times A_{f,i} \text{ s}^{-1}$. The significant quantity in the transition rate $A_{f,i}$ is the factor $w_{if}^3 p_{if}^2$. For $w = 1.0 \times 10^{12} \text{ rad s}^{-1}$ and provided $\Delta n \ll n$, the corresponding change in principal quantum number (from equation (2)) is $\Delta n = \hbar^3 n^3 w_{if} / (mG^2 m^2 M^2) \sim 3 \times 10^{28}$ if for $n = 1.3 \times 10^{34}$ (proton-occupied eigenstates) and $\Delta n \sim 10^{26}$ for $n = 7.7 \times 10^{30}$, (electron-occupied eigenstates). In $A_{f,i}$ the frequency of radiation has increased by a factor of around $10^{30}$ compared to that for a transition with $\Delta n = \pm 1$, but from table I $p_{if}$ has decreased by a far greater amount, so that the product $w_{if}^3 p_{if}^2$ is still very much reduced. It is clear that under these conditions, promotion of a state $(n_i, l_i) = (n_i, n_i - 1)$ to a state $(n_f, l_f) = (n_i + 10^{26}, n_i)$ is impossible in the present lifetime of the universe. For transitions originating from states with $l \neq n - 1$ (i.e. $p > 1$), similar orders of magnitude for $p_{if}$ should apply, provided that $p(= n - l) \ll n$, that is provided $l \sim n$. Similar results will thus almost certainly apply for initial states on any of the first million diagonals of Fig. 1 at least. Significant alteration to the high $n, l$ eigenstates at the present epoch due to stimulated emission/absorption of CMB photons is therefore almost certainly minimal but this does need more detailed quantitative examination. The situation is even less clear for deeper states. This is a very complex problem because of the need to calculate the overlap integrals.

**(b) *Stray galactic radiation***

Stray galactic radiation should not significantly induce stimulated emission/absorption. The derivation of the overlap integral is contingent on the radiation field being uniform over the entire wave function. In scenarios involving traditional atomic transitions the electromagnetic field invariably behaves in this way because the electronic wave functions are generally embedded within the field. Indeed the term $\exp(i\mathbf{k} \cdot \mathbf{r})$ in the transition probability formula $P_{if} = \dfrac{p e^2}{2\hbar^2} |\langle f | \mathbf{r} \cdot \mathbf{E}_w \exp(i\mathbf{k} \cdot \mathbf{r}) | i \rangle|^2$ (Corney[31]) is expanded as a Taylor series on the basis that $\mathbf{r}$ is small over any region where the wave functions are significant

so that that $\mathbf{k} \cdot \mathbf{r} \ll 1$ is much less than unity. With macroscopic eigenfunctions however it is quite feasible that a relatively localised photon wave function may only occupy a small fraction of the region of space occupied by the eigenstate wave functions and the assumption of putting $\mathbf{k} \cdot \mathbf{r} \ll 1$ is not justified. The value of the overlap integral may therefore be considerably different to that calculated with the photon field spread uniformly over the eigenstates. (This is not a problem for CBR because of its homogeneous spread across the entire eigenfunction region.) When considering radiation from the galaxy however, different spatial regions within the particle wave function are exposed to different intensities and spectral qualities of the electromagnetic photon field. The electromagnetic radiation field intensity near a stellar surface may be very great, but not sufficient to induce transitions simply because it interacts with an insignificantly small part of the particle eigenfunction. Nevertheless it is important to attempt to estimate whether galactic radiation could influence macroscopic eigenstructures. This is carried out by assuming a homogeneous density of stray galactic radiation (like the CMB photon field) based on averaging the background galactic radiation across an entire eigenstate. By treating the galactic radiation field as uniform across the eigenstate the assumption is made that $(\mathbf{k} \cdot \mathbf{r})_{average} \ll 1$, which although clearly approximate, should give an estimate of the effects of such radiation.

The average energy output of the Sun is 3.6 x $10^{22}$ W and the luminosity of the galaxy is rated at $10^{10}$ times the solar output. The galaxy is then approximated as a radiating black body of total power output of 3.6 x $10^{32}$ W with a typical stellar surface temperature of 8000 K. Using the relationship that the integrated spectral radiancy is $R = \sigma T^4$ where $\sigma$ is the Stefan-Boltzmann constant and $T$ the temperature, the galactic power corresponds to a black body sphere of radius 8 x $10^{12}$ m. It is assumed that this sphere of radiation is small compared to the radial position of the eigenstate ($r \sim 10^{20}$ m) and hence behaves as an equivalent point source. The integrated energy density $\int_0^\infty \rho(\omega)\,d\omega$ at radius $r \sim 10^{20}$ m is then $10^{32}/(4\pi r^2 c) \sim 10^{-17}$ J m$^{-3}$. Although the spectral quality of the radiation does not alter significantly on its outward journey, the galaxy does not behave as a closed cavity and the intensity, and hence effective energy density, decreases. This effective energy density $\rho_{eff}(\omega)$ across the *entire* eigenfunction is approximated here as

$$\frac{F \hbar \omega^3}{\pi^2 c^3 \left(\exp(\hbar\omega/kT) - 1\right)} \tag{16}$$

where $F$ is considered an 'energy density dilution' factor for the radiation as it expands into the outer halo, to be determined from the total integrated energy density. Performing the integration and taking the beginning of the eigenfunction as $r \sim 10^{19}$ m (ensuring the maximum possible value of $r_{eff}(w)$ over the eigenfunction range), gives $F$ as approximately $8.5 \times 10^{-17}$.

For $T = 8000$ K, the maximum value of $r_{eff}(w) = 3.4 \times 10^{-32}$ J m$^{-3}$s rad$^{-1}$, at $w = 5 \times 10^{15}$ rads$^{-1}$. This angular frequency corresponds to $\Delta n \sim 10^{31} - 10^{32}$ for protons and $\Delta n \sim 10^{29} - 10^{30}$ for electrons. Again the final row of TABLE I demonstrates the overwhelming influence of $p_{if}$ rather than $w_{if}$ on the size of the overlap integral and that such stimulated emission and absorption rates are negligible. As already mentioned, near a high intensity source such as a stellar interior or supernova explosion, where $r(w)$ has the intensity many orders of magnitude larger than those just discussed, the photon density is large only over a relatively very small fraction of the eigenstate volume and it can be estimated that the ability to induce transitions within the eigenstates remains insignificant. For example at the stellar surface $F = 1$, since we now have an approximate black body, but the volume of the star encompasses a volume that is only $\sim 10^{-43} \times$ the volume of the eigenstate.

### (c) *Other Scattering Processes*

This section considers other elastic and inelastic scattering processes. Arguing in a similar way to that of the previous section, equations for describing elastic scattering processes in the atomic domain such as Thomson and Rayleigh scattering are derived on the basis that the field of the photon, (that is, electric field **E** or vector potential **A**) is a plane wave that acts over all of the region where the scattering potential is significant. In deriving an equivalent expression for scattering from the halo eigenstates this is not the case as the scattering centre, the individual particle forming each eigenstate, is subject to a field that is averaged over a much larger region and the probability of scattering reduced accordingly. Essentially this is classically equivalent to saying that a single photon and single particle in a single macroscopic eigenstate have a low scattering probability because they have a low number density in the large volume being considered. From a classical viewpoint therefore it could be argued that the rate of scattering events from a group of photons and particles in eigenstates contained within a given volume is identical to that when the particles are in simple random motion. In that case the present day number density of halo particles is sufficiently low that elastic scattering of photons passing through a halo would not be detectable. It is worth reviewing the problem from the quantum viewpoint however.

Including the possibility that the eigenstate particles can absorb or lose energy (eigenstate excitation or de-excitation) is equivalent to a form of Compton scattering. Rather than describing the process in terms of Compton or inverse Compton scattering off free electrons, the process in instead a type of bound-bound Compton or Raman scattering process: $\hbar w_i + s_i \to \hbar w_f + s_f$ ( $s_i$, $s_f$ the initial and final eigenstates, $\hbar w_i$, $\hbar w_f$ the initial and final photon states. Such 'level resonant' Compton scattering processes have been studied on the atomic scale, for example by Jung[34] who has derived cross sections for Compton scattering allowing for the presence of the $1s \to 2p_0$ electron transition resonance using a screened potential in dense hydrogen plasmas. The differential cross section for bound inelastic Compton scattering is calculable using the lowest order two photon perturbation Hamiltonian $H_I$. Quantum scattering theory gives the differential cross section as $ds_{fi} = \frac{2p}{\hbar c} |\langle f; \mathbf{k}_f \hat{\mathbf{e}}_f | H_I | i; \mathbf{k}_i \hat{\mathbf{e}}_i \rangle|^2 d(E_f - E_i) \frac{d^3 \mathbf{k}_f}{(2p)^3}$ where $|f_i\rangle = |i; \mathbf{k}_i \hat{\mathbf{e}}_i\rangle$ and $|f_f\rangle = |f; \mathbf{k}_f \hat{\mathbf{e}}_f\rangle$ are the initial and final combined atomic state/photon vector systems with $\mathbf{k}_i$ and $\mathbf{k}_f$ the photon wave vectors and $\hat{\mathbf{e}}_i$ and $\hat{\mathbf{e}}_f$ the unit polarisation vectors. Here the overlap integral involves with the vector potential $\mathbf{A}$ of the field and the momentum operator $(\hbar/i)\nabla$. Jung shows that, when photon energy is high and away from the $1s \to 2p_0$ resonance (photon energies near transition resonances have been dealt with in the previous system), terms involving $(\hbar/i)\nabla$ are negligible and the overlap integral becomes $\frac{ds_{fi}}{d\Omega} = \frac{w_f}{w_i} \left(\frac{e^2}{mc^2}\right)^2 |(\hat{\mathbf{e}}_i \cdot \hat{\mathbf{e}}_f)\langle f|e^{i(\mathbf{k}_i - \mathbf{k}_f)\cdot \mathbf{r}}|i\rangle|^2$ where $w_i$ and $w_f$ are the initial and final photon frequencies and $d\Omega$ is the solid angle in the direction $\mathbf{k}_f$. Consider states on or near the $p=1$ diagonal of Fig. 1 ($p \sim 10^7$ or less). These states have limited radial extent and radial position that depends on $n$. In order for the quantity $|\langle f|e^{i(\mathbf{k}_i - \mathbf{k}_f)\cdot \mathbf{r}}|i\rangle|$ not be zero, $|i\rangle$ and $|f\rangle$ must be non zero over some common region. The mathematics of Appendix B shows that to achieve this, $|i\rangle$ and $|f\rangle$ must differ in their $n$ values by no more than $10^{21}$ for $n \sim 10^{34}$. From Fig. 1 it can be seen that, for $p \leq 10^7$, this leads to a maximum change in the eigenstate angular momentum number of $|\Delta l| \leq 10^{21}$. Letting the initial and final angular momenta of the eigenstate particle $E$ and photon $g$ be $\mathbf{L}_{Ei}$, $\mathbf{L}_{Ef}$, $\mathbf{L}_{gi}$ and $\mathbf{L}_{gf}$ respectively this implies that the maximum change in the angular momentum of the eigenstate is $|\Delta \mathbf{L}_E| = |\mathbf{L}_{Ef} - \mathbf{L}_{Ei}| \approx \hbar \Delta l \leq 10^{-13}$ J s. By conservation of momentum, $\Delta \mathbf{L}_g = \mathbf{L}_{gi} - \mathbf{L}_{gf} = \hbar \mathbf{r} \times (\mathbf{k}_i - \mathbf{k}_f) = -\Delta \mathbf{L}_E$. Now provided $\mathbf{r}$ and $\mathbf{k}$ are not 'exactly' perpendicular $|\hbar \mathbf{r} \times (\mathbf{k}_i - \mathbf{k}_f)| \sim \hbar r |\mathbf{k}_i - \mathbf{k}_f| \leq 10^{-13}$ J s or $|\mathbf{k}_i - \mathbf{k}_f| = \Delta k \leq 1$ m$^{-1}$. This means is that $|\langle f|e^{i(\mathbf{k}_i - \mathbf{k}_f)\cdot \mathbf{r}}|i\rangle|$ and the scattering cross section will be negligible unless $\Delta k \leq 1$ m$^{-1}$. More importantly, it means that when $|\mathbf{k}_i| \gg \Delta k$, $|\mathbf{k}_i| \approx |\mathbf{k}_f|$ and that the angle between $\mathbf{k}_i$ and $\mathbf{k}_f$ is necessarily small. Thus there is only potential for scattering of photons from the eigenstate particles when the wavelengths of the photons are substantially greater than 6m (50 MHz). Higher frequency photons will pass through the halo essentially undeflected.

Three further points are noted here. The first is that, as with elastic scattering, the model of bound/bound Compton scattering assumes a uniform plane wave photon field acting over the entirety of the eigenstate volume. This is clearly not the case with the macroscopic eigenstates and it would be a reasonable supposition that the effect of this would be to strongly reduce the rate of scattering because the average amplitude of the photon field is so much smaller. It is again an example where the quantum behaviour of an eigenstate might be considerably different from traditional orbiting 'particle-like', localised wave packets. Secondly, considering the halo as consisting of classical free particles, then even at present day particle number densities, classical considerations of photon scattering yields a low rate. The third point to note is that it has only been shown here that, from conservation laws, short wavelength photons will not scatter at large angles. Nothing has been said about the likelihood of scattering events or scattering angle when the photon wavelength is long, only that it is possible according to conservation laws. The differential scattering cross sections and rates will require detailed analysis involving the angular and radial parts of the overlap integral and will depend on the photon and eigenstate particle densities.

### (d) *Scattering by Particles*

Classical particle-particle scattering at present day halo particle densities ($\sim 10^6 \, \text{m}^{-3}$ or less) might be expected to be negligible. A quantum mechanical treatment of particle scattering off occupied macroscopic eigenstates, although complex, should be able to be carried out by replacement of the photon field Hamiltonian used by Jung, with suitable interaction potentials associated with the scattering particles (e.g. coulomb field, Morse field etc.). The overlap integral will then be of the type $|\langle f|H|i\rangle|$. The previously considered states that exhibited low spontaneous transition rates would again be expected to behave in a similar way since there is a limited amount that any operator $H$ can do to affect the magnitude of $|\langle f|H|i\rangle|$. As already mentioned whenever $|i\rangle$ and $|f\rangle$ have substantial energy difference they either share no common region of space or have sufficiently different spatial oscillation frequencies that any overlap integral is effectively cancelled. Whilst operators may modify $|i\rangle$ it is difficult to see how it could be done in a manner so as to render the quantity $|\langle f|H|i\rangle|$ to be large when $\langle f|i\rangle$ is essentially zero.

**F. Radial density profiles**

The radial density profile of dark matter formed from macroscopic eigenstates will depend on the available number of states and the mass of the particles occupying those states. In a sparsely filled structure (that does not collapse because of the long radiative lifetime of its occupants) the density is almost certainly also a function of formation history. In the evolutionary scenario discussed in the next section it is suggested that there will be a preferential transfer of particles to the high $n$, $l$ value eigenstates that have the longest lifetimes. It is of interest therefore to examine what radial density profile would result if the high $n$, $l$ eigenstate array was either completely filled or partially but uniformly filled over all the available states. The earlier discussion of transition rate probabilities suggests that the state lifetime depends on the average on the value of $p(\equiv n-l)$, with the longest lived states on the $p=1$ diagonal. What is required therefore is to calculate the density profile of matter formed by the occupation of a trapezium-shaped strip of states in Fig. 1, defined by a horizontal line $n = n_{max}$ representing a maximum $n$ value and the diagonal lines $p=1$ and $p = p_{max}$, where $p_{max}$ is a diagonal whose states still have a sufficiently long lifetime to be considered stable. In this array of states $l$ runs from $l = n-1$ to a minimum value $l = n-1-p_{max}$ and for each $l$ there are $2l+1$ $m$ values. Ignoring spin, the total number of states $N$ up to $n = n_{max}$ is

$$N = \sum_{i=1}^{p_{max}} \left( \sum_{p=1}^{i} (2(i-p)+1) \right) + \sum_{i=p_{max}+1}^{n_{max}} \left( \sum_{p=1}^{p_{max}} (2(i-p)+1) \right) \quad (17)$$
$$\approx p_{max} n_{max}^2$$

if $p_{max} \ll n_{max}$.

For any value of $l$ (or $p$) the sum over all $m$ states gives a spherically symmetric density distribution. Now $\rho(r) = dM/(4\pi r^2 dr)$. The outer halo regions being considered here involve states that have $p \ll n$ and so that, relative to their radial position, their radial extent is small. This means for any eigenstate with quantum number $n$, the relation $r = n^2 b_0$ applies so that $dr = 2 n\, b_0\, dn$. If $n_{max}$ is allowed to increase by $dn_{max}$, then all the states (of number $dN$) between $n_{max}$ and $dn_{max}$ will all lie between $r$ and $r + dr$. Differentiating equation (17) with respect to $n_{max}$ and noting that $dM = mdN$, where $m$ is the mass of the eigenstate particle, gives

$$\rho(r) = \frac{dM}{4\pi r^2 dr} = \frac{mdN}{4\pi r^2 dr} = \frac{2m\, p_{max} n_{max} dn_{max}}{4\pi r^2 dr} = \frac{m\, p_{max}}{4\pi r^2 b_0} \quad (18)$$

That is the density of matter in the eigenstructure is proportional to $1/r^2$ in the outer parts of the halo. If the eigenstates that have lower $n$ values are not populated as densely (because the lower $n$ valued eigenstates have shorter lifetimes) then the radial profile will be shallower, perhaps even flat, at lower radii)

## IV. FORMATION AND EVOLUTIONARY SCENARIO

From the previous sections it might seem that, because of their limited range and close energy spacing, interaction of macroscopic eigenstates with 'normal' matter would prove difficult under any circumstances. If this is so then it is hard to imagine how eigenstates and eigenstructures might ever form. A mechanism by which baryonic particles can be transferred into eigenstates to produce an eigenstructure and a hypothetical evolutionary development of the eigenstructure halo is required. This section examines a scenario that has the potential to explain the formation and evolution of dark matter halos. In this scenario, eigenstructures form around the seed potentials of supermassive primordial black holes. Particles undergo a type of 'gravitational recombination' (or more accurately 'gravitational combination' since, as in primordial $p^+/e^-$ recombination, the particles were not together to start with) into the gravitational energy levels of the eigenstructure analogous to the normal recombination of ions and electrons in a cooling plasma. A fundamental difference however is that in the formation of dark matter halos, the 'cosmic plasma' is rapidly expanding at the same time as it is cooling. Unlike a $p^+/e^-$ recombining plasma, this results in particles remaining uncombined, even though the average temperature has fallen well below that required for recombination, simply because the density has fallen so quickly that the rate of recombination has dropped to an insignificant level before the gravitational recombination process has completed. The hypothesis is that the hotter, uncombined particles at the high energy end of the velocity distribution, and the particles in the volumes in between the potential wells of the eigenstructures, form the traditional baryonic matter observed at the present day.

### A. Formation of primordial black holes

It is well known that primordial black holes (PBH) can form from over-density regions at the various phase transitions in the early universe (Carr[35,36], Jedamzik[37], and Afshordi, McDonald, & Spergel[38]). At phase transitions black holes might be expected to form with masses up to the horizon mass. This gives an expected upper limit on PBH masses during radiation dominated eras as

$$M_{PBH} = M_H(T) = 1 M_{Sun} \left( \frac{T}{100 \text{MeV}} \right)^{-2} \left( \frac{g_{eff}}{10.75} \right)^{-\frac{1}{2}} \qquad (19)$$

where $T$ is the temperature, $g_{eff}$ is the number of degrees of freedom and $M_{Sun}$ is the solar mass. Furthermore numerical calculations by Hawke and Stewart[39] suggest a lower limit to $M_{PBH}$ as $\sim 10^{-4} M_H$. The PBHs required for eigenstructure formation need to be massive and would have formed at the $e^+/e^-$ phase transition at $t \approx 0.75$ s. Taking $g_{eff}$ as $(43/4)$ (Olive and Peacock[40]) gives $M_H \sim 10^5 M_{Sun}$. Afshordi *et al*[38], in considering black holes as dark matter candidates, allow for PBH masses up to $10^6 M_{Sun}$. It therefore might be expected that at this phase transition PBHs were formed predominantly in the range of masses from $10^{31}$ to $10^{35}$ kg and possibly as high as $10^{36}$ kg. Hall and Hsu[41] use cosmological considerations to fix a maximum limit on the number of PBHs as one per $10^7$ horizon volumes whilst the work of Carr[35] provides a formula for the number density of black holes per gram of mass. Hypothetically, each black hole seed potential well, if it is sufficiently deep, goes on to form an eigenstructure dark matter halo, which in turn forms a framework for the formation of galaxies and/or galactic clusters. If the number of galaxies at present is estimated as $2 \times 10^{11}$ per present Hubble volume, this extrapolates back to only one PBH per $\sim 10^{13}$ horizons at the formation time of $t \approx 0.75$ s, a figure which may be more in keeping with the expected number of over-density regions and the work of Carr (see section C below). It may be that there are more, smaller galaxies in the present universe than have been so far observed. It is possible that every centre of visible accumulation of material, even down to the globular cluster scale, contains a central black hole and possibly also a small halo.

## B. Development of the Eigenstructure

To make estimates about the rate of population of an eigenstructure halo is necessary to know the radiation density $\rho_{rad}(t)$, particle density $\rho_{mat}(t)$ and temperature $T$ from $t \approx 0.75$ s onwards, as well as estimates of the quantum 'gravitational recombination' rate. Temperatures were calculated from $T = (1/k_B)\sqrt{32\pi^3 G\, g_{eff}/(90\hbar^3 c^5)}$. Letting $\rho_{rad}(t)$ and $\rho_{mat}(t)$ represent densities both before and after the matter/radiation equivalence time $t_e$, gives $\rho_{rad}(t)(\text{pre-}t_e) = 3c^2/(32\pi G t^2)$ and $\rho_{mat}(t) \propto t^{-\frac{3}{2}}$. Post-$t_e$, $\rho_{mat}(t) \propto t^{-2}$ and $\rho_{rad}(t) \propto t^{-\frac{8}{3}}$. Knowing the present radiation density (including neutrinos) enables the constants of proportionality to be obtained and hence values for the matter and radiation densities at all relevant times. (As a check on accuracy, the value of present matter density $\Omega_m$ was calculated using this method as 0.253, compared to the WMAP figure of $0.27 \pm 0.04$.) At $t \approx 0.75$ s, the particle and radiation mass densities are $\sim 8 \times 10^8 \,\text{kg m}^{-3}$ and $\sim 4 \times 10^3 \,\text{kg m}^{-3}$ respectively. The temperature of $\sim 10^{10}$ K corresponds to a mean thermal energy of $\sim 10^6$ eV.

As already mentioned, particles populate eigenstates by the process of gravitational recombination. In atomic recombination, the equilibrium fractions of recombined/ionised particles are given by the Saha equation, which is given, in a simple case as

$$\frac{n_e n_i}{n_a (n_a - n_i)} = \frac{2\pi m_e k T}{\hbar^3} \exp\left(-\frac{E_i}{kT}\right) \qquad (20)$$

where $n_i, n_e$ and $n_a$ are the ion, electron and atom densities and $E_i$ the ionisation energy of the level concerned. In the analogous gravitational case, an eigenstate would not be populated until the thermal energy of the particle falls to (or below) the eigenstate's gravitational equivalent 'ionisation' energy. This means that only eigenstates that have an energy level value that is substantially deeper than the thermal energy will be available for population. In the atomic case the population of bound states depends on the recombination rate, and the collisional and photo ionisation rates. Particles transferring to shallower gravitational states will be excited and ultimately removed by radiation and collisions at a rate which may be as fast as they are forming, in an analogous way to the atomic situation (i.e. a process of "gravitational ionisation"). From the Saha equation, it might be expected that all matter should be in the form of eigenstates because the present day temperature is so low. However it is worth restating that the continually declining densities in the universe mean that the gravitational recombination rates have dropped to such an extent that equilibrium is never achieved and some fraction of traditional matter remains uncombined into eigenstates, this gravitationally uncombined matter being the baryonic matter we detect today.

The rate of 'gravitational recombination' will depend on the transition probability, the density of particles and the density of eigenstates available. The quantum mechanical treatment of the equivalent atomic recombination rate in hydrogen and helium has been examined by Hirata[42] and Menzel and Pekeris[43] and shown to be

$$R_{rec}(n,l,m) = \frac{4 e^2 R\, n_p}{3 \boldsymbol{p}\, c^3 \hbar^4}\left(\frac{\hbar^2}{2\boldsymbol{p}\, m_e k_B T}\right)^{3/2} \int_0^\infty dk \sum_{l,m} E_g^3 (1+\bar{N}) |\langle k\, l\, m|\mathbf{r}|n\, l\, m\rangle| \exp\left(-\frac{\hbar^2 k^2}{2 m_e k_B T}\right)$$

(21)

where the $|k\, l\, m\rangle$ are the 'unbound' particle wavefunctions with quantum number $k$, normalised over a large radius $R$, $|n\, l\, m\rangle$ are the 'bound' states, $E_g$ is the energy of the photon emitted during the recombination process, $\bar{N}$ is the average number of photons per state with energy around $E_g$, $n_p$ is the particle density and the other symbols have their usual meanings. This is a difficult quantity to recalculate in the gravitational case since it involves the sum of overlap integrals over many states with varying degrees of overlap. In the present introductory approach the assumption is made that the rate of recombination shortly after PBH formation is rapid since the average density of particles at $t \approx 0.75$ s is $\sim 10^{25}$ greater than today ($10^{29}$ m$^{-3}$ compared to $10^5$ m$^{-3}$) with a collective wavefunction comprising a particle density that extends throughout the entire volume of interest. The assumption is that the gravitational recombination rate is sufficiently rapid at early times to allow eigenstate formation, and this assumption is yet to be verified via a rigorous quantum mechanical treatment. Under this assumption the rate of formation of the eigenstructure is determined by the rate at which eigenstates become thermally available for population. Furthermore, as the universe expands, a simple Newtonian approximation is assumed to be applicable where the eigenstate energy is determined by the effective central mass that the eigenstate 'sees'. This in turn depends on the relative density of the matter inside the eigenstate radius relative to the density of the surrounding universe.

It is possible to develop a simplistic model to represent the rate of filling of the eigenstructure. Using the result from the Saha equation that the thermal background energy needs to fall below a certain threshold before recombination can occur effectively, one can introduce the concept of a 'thermal radius' $r_{therm}$. The thermal radius is defined as the radial distance from the centre of the structure where the depth of the potential well is such that eigenstates lying within this radius are at energy levels that are of the order of the background thermal energy or deeper. Eigenstates that lie predominantly within this spherical volume are capable of being populated at the given time via gravitational recombination. The thermal radius will be larger the lower the thermal background temperature and the deeper the potential well.

Clearly the thermal radius increases in size with time because of two reasons. Firstly, the average thermal background energy is decreasing because the universe is cooling. Secondly, as particles are added to the eigenstructure, its effective mass increases due to this capture of matter, and to surrounding universal expansion beyond the forming eigenstructure. It is assumed in this Newtonian scenario that the potential well in the vicinity of the eigenstate formation is effectively deepened by these processes. A further important assumption is that, because the well deepening is formed by the universe expanding 'away from' the eigenstructure, it does not take the normal relativistic time for particles to feel the changing potential - it is produced concurrently everywhere with the expansion. To preserve the BBN elemental abundance ratios, the bulk of the eigenstructure needs to form before the end of the BBN era. Whilst this is possible in the above scenario, there is at least one potentially serious problem to overcome. At the time of formation of the PBH it has been assumed that, at least outside the over-density region, its potential well of the PBH is pre-existing, since it was formed from a pre-existing over-density region. This may be possible but equation (25) yields a thermal radius that for some time travels faster than the light horizon. Whilst not the movement of a physical object, it involves the notion that the increase in the depth of the potential well experienced by particles on the verge of the thermal radius increases because of the ongoing decrease in the surrounding density of the universe. The possibility that this can happen remains to be investigated and verified. One hypothesis to circumventing this potential problem relates to the temperature of the background. As the eigenstructure forms, the slower particles are captured and more radiation is emitted. This could have a significant effect on keeping the temperature high for a longer period which may (1) delay the BBN process and (2) cause sufficient inhomogeneity to require recalculation of the BBN elemental ratios. If this were true, it may be possible to develop further scenarios where the structure could take much longer to grow and still not conflict with BBN.

Returning to the concept of the thermal radius, it is possible using the simplistic model to obtain the growth of the eigenstructure halo with time, the maximum mass capacity of the halo and ultimately make an estimate of the ratio for the expected amount of baryonic to total matter. By equating the thermal background energy $(kT)$ with a given fraction $f$ of the eigenstate ionisation energy, a value of the quantum number $n = n_{therm}(t)$ is obtained such that for all states $n \leq n_{therm}(t)$ at time $t$, the 'gravitational ionisation' energy $E_n$ of the eigenstate, will be greater than $kT/f$, the energy depth required for effective gravitational recombination. A value of $n_{therm}(t)$ can be related to the thermal radius $r_{therm}$ because the high $n$ states have a limited range out to $r_{max} \sim 2n^2 b_0$. Initially the thermal radius is determined by the mass of the primordial black hole $M_{PBH}$ and the temperature at $t \approx 0.75$ s. As the universe cools and expands $r_{therm}$ successively increases and so does the effective central mass as seen by particles at $r = r_{therm}(t)$. As time continues, particles are deposited into the eigenstates in successive shells of gradually reducing density. The effective mass at any time $t$ is the sum of the PBH mass plus the contribution of the effective masses from each of the shells of thickness $\Delta r_{therm}$. The following relations can be obtained:

$$n_{therm}(t) = \frac{G m M_{eff}(t)}{\hbar} \sqrt{\frac{m}{2 f k_B T}} = k_n M_{eff}(t) T^{1/2}$$

$$r_{therm}(t) = 2\, n_{therm}(t)^2 b_0$$

$$T(t) = \frac{1}{k_B}\left(\frac{90\, \hbar^3 c^5}{32\, \pi^3 G\, g_{eff} t^2}\right)^{1/4} = k_T\, t^{-1/2} \qquad (22)$$

$$\rho(t) = k_m\, t^{-3/2}$$

$$M_{eff}(t) = M_{PBH} + \sum_i \Delta M_i(t_i, t) = M_{PBH} + \sum_i \left(4\pi\, r_{therm}(t_i)^2\, \Delta r_{therm}(t_i)\right)\left(\rho(t_i) - \rho(t)\right)$$

In the above equations, $M_{eff}(t)$ is the effective central mass as seen by a particle just outside the $r_{therm}(t)$ shell (of thickness $\Delta r_{therm}$), $\Delta M_i(t_i, t)$ is the effective mass contribution that the $i^{th}$ shell, formed at $t_i$, makes at time $t$, $\rho(t)$ is the matter density of the universe at time $t$ and $k_n, k_m$ and $k_T$ are constants. The following simple differential equation relating $r_{therm}(t)$ to $t$ can then be obtained:

$$\frac{\partial r_{therm}(t)}{\partial t} - k_1 \frac{r_{therm}(t)^3}{t^2} - \frac{r_{therm}(t)}{2t} = 0 \qquad (23)$$

where $k_1 = 4\pi k_m k_n \hbar^2 / (mGm k_T)$ is another constant. Initial values of the relevant quantities are defined as:

$$T(t_0) = k_T t_0^{-1/2}, \; n_{therm}(t_0) = \frac{k_n t_0^{1/4}}{k_T}, \; r(t_0) = k_m t_0^{-3/2} \text{ and } r_{therm}(t_0) = \frac{2b_0 k_n t_0^{1/2}}{k_T} \qquad (24)$$

Equation (23) has solution

$$r(t) = \frac{r_0 \sqrt{t/t_0}}{\sqrt{1 - 2\left(\frac{k_1 r_0^2}{t_0}\right) \ln\left(\frac{t}{t_0}\right)}} \qquad (25)$$

If $f \sim 1$, this equation shows that it is thermally possible for an eigenstructure of $2 \times 10^{42}$ kg to form within a few seconds provided the initial PBH seed mass is about $1.2 \times 10^{34}$ kg or greater (and provided the gravitational recombination rate is rapid). Even if $f \sim 1/10$ (although Saha equation applied atomic systems might suggest this figure, it would not necessarily be this small), $2 \times 10^{42}$ kg eigenstructures will form in a similar time provided that the PBH seed mass is $\sim 4 \times 10^{35}$ kg or greater. All PBHs above a certain size $(\sim 10^{34} - 10^{35}$ kg$)$ should form large galactic halos, the growth limited only by the presence of neighbouring wells. If the PBH mass is smaller than these critical values smaller eigenstructures will be formed but possibly engulfed by the spread of the larger structures and take longer to grow. Black holes with still smaller masses will take so long to populate their eigenstates that they will hardly form halos at all before the universal matter has expanded away from them. It is nevertheless assumed that there will be various sized eigenstructures because of the various sized PBHs that were formed at the phase transition. However because the range of masses for effective formation of large halos has such a sharply defined cut-off, it means that the size of the smaller structures is determined primarily by the increased formation time. Because the larger eigenstructures can potentially form in a time before the onset of big bang nucleosynthesis (BBN), it means that elemental abundances may not be affected. At the very least BBN ratios would need to be recalculated in the context of the eigenstructure scenario, since the particles left uncombined are hot and the reaction rates with eigenstate particles may be different. Since different mass eigenstructures take different times to form, the smallest ones may be still forming during the BBN era. This means that there is a possibility that the elemental ratios of the various eigenstate particles from BBN may be different in different in smaller sized halos.

At the time of formation of the eigenstructure, its outer radius is only $\sim 10^{13}$ m and from equations (22) to (25), the initial radial density profile $r(r,t_0)$ would not be expected to be $\propto 1/r^2$. (Also from equations (22) to (25) it can be shown that, $r(r,t_0) \propto 1/r^n$, where $n \sim 0$ for $r < r_0$, $n \sim 0.4$ to 3 when $r$ is just greater than $r_0$ (depending on $M_{PBH}$), and that $n \to 0$ again as $r$ becomes larger than $r_0$.) At the time of eigenstructure formation the radiation energy density was $\sim 10^{26}$ J m$^{-3}$, with a photon number density of $\sim 10^{37}$ m$^{-3}$ existing over the entire eigenstructure. The ratio of stimulated to spontaneous transitions is $c^3 \mathbf{p}^2 g_f \mathbf{r}(\mathbf{w}) / (\hbar g_i \mathbf{w}_{if}^3)$. This shows that the stimulated rates are much higher when $\mathbf{r}(\mathbf{w})$ is large and $\mathbf{w}_{if}$ is small. Additionally, across the total regime of $l$ values there will be many pairs of states whose overlap is substantial enough for the states to participate in stimulated transitions both up and down. Even at the extremely low frequency end ($10^{-17}$ Hz) of the high temperature radiation spectrum, the energy density of photons at that corresponds to transitions of the type $A \to A'$ of Fig. 1 is $10^{10}$ times higher than that for similar frequency photons in the present day CMB. The situation is more complicated than this however. During and shortly after the initial eigenstructure formation, there are many things different to that in the present day halo. The effective mass, the value of the parameter $b_0$, the value of $n$, the radiation density, the frequency of the transitions etc. are all very different from their present day values. When these factors are taken into account, for particles at or near $r_{thermal}$ (which correspond at this time to $n$ values of $\sim 10^{25}$), it is possible to calculate the stimulated transition rate. An $A \to A'$ type transition in a typical scenario based on a PBH mass of $1.4 \times 10^{34}$ kg has a stimulated transition rate of $2 \times 10^{15}$ s$^{-1}$ at the end of the formation time (compared to $500$ s$^{-1}$ for the 'halo edge' particles today). This is not sufficient to expand the halo via $A \to A'$ type transitions. Likewise, because of equation (59), direct transfer from a low $l$, high $n$ state to is not possible. However as time proceeds (between 14 and 250 seconds) conditions change in such a way that, for the example considered, the $A \to A'$ type transition rate actually rises to $10^{34}$ s$^{-1}$. This pushes the eigenstates out to higher $n$ values but the rate drops rapidly as $n$ increases. For each $n$ value there is a time that corresponds to a maximum transition rate, but that rate decreases with increasing $n$. When $n = 10^{29}$ that rate is about $10^{23}$ s$^{-1}$ at a time of about 50 s. Since all these calculations are relevant only to $A \to A'$ type transitions, further investigation is needed to look at other transitions. 'Deeper' (low $l$ states) have wider radial spread and potentially more overlap, and less dramatic differences in the spatial radial oscillation frequencies for different $n$ values, therefore providing greater potential for possible transition resonances.

Another alternative is that the transition into eigenstates does not happen as rapidly as been assumed. The structure may still gravitationally retain the particles but they transfer to eigenstates at progressively later times. Although the radiation density is falling transitions can occur to higher $n$ values at later times and radiation may have a lesser part to play in the formation. The basic hypothesis is however that that over time, the radiation field mixes the eigenstate population and a net transition of eigenstates takes place toward high $n, l$ eigenstates through the combination of stimulated transitions and spontaneous decay. It is reasonable to suppose that this process could occur to a sufficient extent to expand the eigenstructure enough over time.

Whatever way it might happen, eventually in this scenario the photon number density has dropped sufficiently to render the stimulated transition rates small, and particles have found their way into a strip of sufficiently long lived, high $n, l$ states that should yield $1/r^2$ density profiles, with the remaining 'uncombined' baryons forming the visible matter of the universe. It is to be expected that the recombination and evolutionary processes continued for a long time and are still continuing today although at a much reduced, insignificant, rates. It is also interesting to note that the size of the average large halo $\left( r \sim 6 \times 10^{21} \, \text{m} \right)$ is the same as the separation of the large galaxies at the time of reionisation ($t = 180$ million years, $z = 20$) which is perhaps the last opportunity for significant halo photon interaction. Further detailed quantitative analysis needs to be carried out on the evolutionary scenario.

It is also important to establish whether the interaction with the radiation field would have had any effect on the CMB. The interactions at times around decoupling and later involve frequencies that are below those examined in the CMB range. For $n$ ranging between $10^{28}$ and $10^{32}$, the wavelengths involved in stimulated interactions at decoupling time are between $\sim 10^7$ and $10^{19}$ m. It may be that CMB will show some irregularities in this range of wavelengths.

**C. Expected number of Eigenstructures**

At the time of formation, PBHs are expected to be randomly distributed throughout the universe. The growth of the halo eigenstructure cannot proceed unchecked. Once the thermal radii of neighbouring eigenstructures meet $(\sim 1 \text{s to } \sim 1000 \text{s})$ the halos are no longer able to grow. Thus the picture of the universe shortly after eigenstructure formation is one that is full of predominantly spherical halos touching at their extremities with a small amount of hot unbound or semi-bound particles. This picture fixes the mass of halos. Taking the number of large galaxies as $\sim 10$ billion, the average spacing between galaxies at the formation time $(\sim 1 \text{s})$ can be calculated at around $\sim 10^{13}$ m. Using equation (25) in combination with equations (22) then predicts that the halo masses of the larger galaxies should be about $2 \times 10^{42}$ kg which is consistent with observations.

The number of PBHs produced may be estimated from the work of Carr[35] who gives an equation for the number density per gram of PBHs produced as

$$\frac{dn}{dM} = (a-2)\left(\frac{M}{10^{15}}\right)^{-a} \left(10^{15}\right)^{-2} \Omega_{PBH}\, r_{crit} \qquad (26)$$

where $a = (1+3g)/(1+g)+1$, $g$ is the constant in the equation of state, $\Omega_{PBH}$ is the fraction of the critical density existing as PBHs and $r_{crit}$ is the critical density. During the radiation era $g = 1/3$ and taking $\Omega_{PBH}$ as $10^{-7}$ and integrating equation (26) gives the density of black holes whose greater than $10^{34}$ kg as $10^{-79}$ m$^{-3}$. This is too low as the density of large galaxies can be estimated at between $5 \times 10^{-70}$ and $10^{-69}$ m$^{-3}$. Of course $\Omega_{PBH}$ may be higher but it is unlikely to be greater than $10^{-4}$ which still only results in a relevant PBH density of $3 \times 10^{-76}$ m$^{-3}$. However Carr[35] also suggests that phase transitions lead to the equation of state becoming soft $g \ll 1$ for a time. If $g \sim 1/100$ (which appears not to be unreasonable), then the density of PBHs with the required mass becomes $\sim 7 \times 10^{-70}$ m$^{-3}$ for $\Omega_{PBH} = 10^{-7}$ (and $\sim 7 \times 10^{-67}$ m$^{-3}$ for $\Omega_{PBH} = 10^{-4}$) which is in much better agreement with the estimated number of large galaxies. Another point about Carr's equation is that it predicts more, smaller PBHs. Isolated small PBHs would not have accumulated very much matter in the early universe either by accretion or in the form of eigenstructures. If formed within the vicinity of a larger halo however, some of these smaller PBHs may have accumulated more matter and acted as seeding structures for globular clusters since the density remains high for long times within the larger halos.

**D. Amount of baryonic matter**

The aspect ratio of the halos may depend on the shape of the original density distribution and the angular momentum of the material in the volume around the original PBH, but it would be expected that most should be spherical. Thus, as a rough approximation, not long after formation, the universe might be pictured as an assembly of spherical halos filling space and touching each other at their extremities. If the radiation has been sufficiently dense, and the induced transition processes rapid enough, then the radial density profiles will be forced into a $1/r^2$ form (see section F. Radial density profiles, earlier in this chapter). The highest and most loosely bound or semi-bound particles should end up promoted into or occupying the 'corners' of the cubical volumes in between the spherical halos. It is interesting (although perhaps just coincidental!) that using a typical $1/r^2$ dark matter density profile (such as that described by Munyaneza and Biermann[33]) gives the fraction of 'non-eigenstate' matter in these corners (i.e. the fraction that, in this scenario, would end up as traditional visible baryonic matter) as $\sim 0.18$. The results of WMAP give the baryon to total matter ratio $\Omega_b/\Omega_m$ as $0.17 \pm 0.01$ [44].

## E. WMAP and Eigenstructures

This picture of the early universe as a 'two fluid' assembly consisting of massive eigenstructures embedded in a hot photon-'visible baryon' background appears consistent with the WMAP results. The eigenstructures before decoupling behave essentially as massive particles. Fluctuations in their number density on large scales reflect the large scale density fluctuations of inflation, the same as in a scenario using traditional WIMPs as the CDM particles. In this sense the development of the large scale structure of the universe, so well predicted by the CDM model is preserved and it is only on the galactic scale that the picture differs significantly. The acoustic peaks in the harmonic power spectrum from WMAP $(l < 1000)$ reflect fluctuations at decoupling on scales that are of super cluster size ($\sim 2 \times 10^{21}$ m at decoupling, $\sim 2 \times 10^{24}$ m today). Fluctuations on the galactic halo scale (i.e. the eigenstructure scale) correspond to angular scales of $\sim 5 \times 10^{-5}$ degrees (i.e. $l$ values $> 2 \times 10^4$) and are therefore well beyond the present region examined by WMAP. Searching for fluctuations at $l \sim 2 - 3 \times 10^4$ may be a way to look for evidence of the existence of eigenstructures in the early universe.

## F. High Energy Cosmic Rays

One significant problem in astronomy concerns the observation of small numbers of ultra-high-energy cosmic rays. At present, no known physical process can account for the very high energies of the particles. The production of such cosmic rays follows quite naturally however from level decay within eigenstructures. In traditional atomic physics the energy release accompanying electron demotion within the atom is of the order of tens of electron volts or less and electromagnetic radiation is the only possible type of emission. In the spontaneous decay of nuclei, however energy changes are much higher and the process of pair production accompanying spontaneous decay is also possible. Low $l$ valued eigenstates have broad radial and angular probability density profiles and although transitions are likely to be rare, resonances may occur. The energy differences between some of the gravitational eigenstates are extremely large and potentially provide vast amounts of surplus energy to be carried away as kinetic energy of the particles of the pair-production process accompanying level decay. It might be expected that such pair production could be observed if the lifetimes of these inner states are sufficiently short. Whilst the bulk of such transitions would probably have taken place some time ago in the Milky Way, such decays may still be continuing at much reduced rates and higher production rates may be observable in young objects such as quasars.

## V. DETECTING EIGENSTRUCTURES

What observations or tests might help verify the eigenstructure hypothesis? If there are observational clues to the existence of eigenstructures then the best place to look is in the long wavelength range of the electromagnetic spectrum. It may be possible to detect ELF radio noise directly from the galaxy (although the rate is likely to be extremely small). More likely is the possibility of detecting enhanced amounts of radiation in the CBM spectrum at wavelengths much longer than is measured at present. The stimulated transitions and decay of eigenstates that is predicted by the evolutionary scenario may have existed up to and possibly post decoupling. If so a deviation from the CBM black body spectrum at the very long wavelengths might be anticipated. Already mentioned is the direct detection of the imprint that eigenstructures themselves might have left on the CMB, observable as peaks in the CMB anisotropy spectrum at high $l$ values $\left(2-3\times10^4\right)$.

Another possible place that might reveal the history of eigenstructure development is in the measurements of BBN ratios. Smaller galaxies that have remained isolated from larger clusters since their formation might show different BBN elemental ratios if they took longer to form. Yet another possibility is to look for evidence of electron-positron or proton-antiproton pair production in galaxy. This consequence of direct eigenstate decay will probably have a rate that is likely to be small. Such events might be more likely to be observed in younger eigenstructures, possibly for example quasars.

It might be expect that most halos would be spherical, particularly if they evolve through stimulated transitions. Deviations from sphericity could be expected to result from significant angular momentum of the original density regions. Stimulated absorption and emission processes thought to mix and expand the eigenfunctions may tend to make halos more spherical. Further investigation of this is required but the finding that the majority of halos were significantly oblate or prolate might be bad news for the eigenstructure hypothesis.

## VI. SUMMARY AND DISCUSSION

The eigenstructure model of dark matter represents a fundamentally different way of picturing matter on astronomical scales. Evidence from atomic and nuclear systems is that bound matter preferentially exists in eigenstates on small scales, because it is only in these states that matter is radiatively stable. There is no reason to believe that matter on large scales will not show the same preference. However, on the macroscopic and astronomical scale at present day particle densities, the rate at which particles collapse into stationary states has been seen to be extremely slow. In the scenario being presented here, it is precisely because of this slow rate of collapse that traditional visible baryonic matter still remains today, uncombined into eigenstructures.

The salient points from the present approach may be summarised as follows:
- Although there is nothing in quantum theory that forbids the existence of eigenstructures, it is a significant leap to contemplate the existence of astronomically sized stationary structures that have been hitherto observable only on the atomic scale.
- There is no need to introduce new particles
- There is no need to introduce new physics
- Concerning halo size, minimum eigenstate binding energy is coincident with the expected halo radius.
- The total halo mass is consistent with a range of masses of those from electrons to baryons. Particles that are much lighter than electrons will not be able to form stably bound halos of sufficient mass and if particles significantly more massive than protons were involved more massive halos (than the observations of galaxies and clusters suggest) might be expected.
- There is a sufficient number of suitably long lived eigenstates to accommodate the mass of the halo.
- For the relevant quantum eigenstates in the significant parts of the halo, radiative lifetimes are such as to render the states sufficiently long lived.
- Long radiative lifetime leads to an inability to undergo gravitational collapse.
- Present day eigenstructures do not emit significant quantities of electromagnetic radiation.
- Present day eigenstructures do not interact significantly with particles or radiation.
- Critical to the gravitational recombination process is the assumption that 'over-horizon' quantum collapse can occur. As mentioned on page 5 such over horizon quantum effects are commonly observed in entanglement and super-luminal quantum connectivity experiments.

- The mass and number density of halos produced in the formation scenario is consistent with the range of number density of suitably sized PBH's.
- The size of halos is the same as the average galaxy separation at reionisation – perhaps the last chance for significant photon halo interaction
- Consideration of the population of what seem to be the longest lived and most stable states, leads to $1/r^2$ density profiles in the outer regions of halos, consistent with observation.
- Lower values of the quantum number $n$ lead to decay times which are lower and suggest that profiles should 'turn over' to form shallower power law structures at lower halo radii. That is there is a strong possibility of explaining the observations of non-cusp like halo profiles.
- Consideration of the space filling sphere model with $1/r^2$ density profiles leads to a prediction that visible to total matter ratio will be about 0.18.
- There is the possibility that eigenstate decay may help explain origin of the ultra high energy cosmic ray events and may also contribute to the energy output of quasars.
- The evolution of eigenstructures suggests that there could be excess quantities of electromagnetic radiation in the CMB at the extremely long wavelength end, into and beyond the ULF bands.
- Eigenstructures should have left an imprint of acoustic peaks at the very high $l$ end of the WMAP anisotropy power spectrum.

Although the models used in this present introductory approach have involved very simplified potentials and large approximations, the broad conclusions from these models should remain generally valid when more realistic details are included. The eigenstructure hypothesis appears to be consistent with most observations. This consistency justifies re-examination of Big Bang models, and models of galaxy formation and evolution, with the inclusion of macroscopic quantum stationary eigenstates at all times in the past. There is certainly much to do in terms of understanding their theoretical development and investigating observationally how the existence of macroscopic gravitational eigenstates and eigenstructures might be tested.

# APPENDIX A. APPROXIMATE EXTENT OF THE FUNCTION $(R_{nl}\, r)^2$ FOR LARGE $n$.

### Case I: $l$ is close to $n$:

The general form of the radial wave functions can be used to obtain a measure of its extent for large $n$. When $l = n-1$ the form of the single peaked wavefunction can be obtained exactly from equation 4 to give

$$(R_{n,n-1}\, r)^2 = \frac{4^n \exp\left(-\dfrac{2r}{n b_0}\right)\left(\dfrac{2r}{n b_0}\right)^{2n}}{n^2 b_0\, r\, (2n-1)!} \tag{27}$$

Solving $\dfrac{\partial}{\partial r}\left[(R_{n,n-1}\, r)^2\right] = 0$ then gives the position of the peak as $r = b_0\left(n^2 - n/2\right)$. An easily calculable measure of the width of $(R_{n,n-1}\, r)^2$ is obtained by introducing a quantity termed here the 'zero concavity width' $\Delta\left[(R_{n,n-1}\, r)^2\right]_{cw}$, defined as the width joining points on either side of the peak where the concavity changes sign. The 'zero concavity width' is still a realistic and relatively accurate measure of the peak width and is obtained by solving $\dfrac{\partial^2}{\partial r^2}\left[(R_{n,n-1}\, r)^2\right] = 0$ to give the two zero concavity points on either side of the curve. The spacing between these points yields a formula for the concavity width of $(R_{n,n-1}\, r)^2$ as $\Delta\left[(R_{n,n-1}\, r)^2\right]_{cw} = b_0 n\sqrt{(2n-1)}$. Similar although increasingly more complex calculations can be made for $(R_{n,n-2}\, r)^2$, $(R_{n,n-3}\, r)^2$, $(R_{n,n-4}\, r)^2$ etc. (It should also be noted that the position of the peak of the function $R_{n,n-1}$ is $r = b_0\left(n^2 - n\right)$ and the width is $\Delta\left[R_{n,n-1}\right]_{cw} = 2 b_0\, n\sqrt{(n-1)}$.)

### Case II: $l \ll n$:

When $n \gg 1$ and $l \ll n$ (i.e. $p \sim n$), the radial probability density function $(R_{n,l}\, r)^2$ has so many terms that it is impossible to write down explicitly. It is possible however to obtain various empirical relationships that enable the properties of these functions to be estimated.

For $n \gg 1$ and $l \ll n$ $(R_{n,l}r)^2$ is a many peaked function or $r$, the final (and largest) peak occurring at a radial position $r_{finalpeak}$. The position of this peak is important not only because it is the largest but also because gives an estimate of the radial extent of $(R_{n,l}r)^2$. Shown in Fig. 2 are plots of $r_{finalpeak}$ versus $n^2$ for $l = 0$, which suggest that $r_{finalpeak} \approx k\,b_0\,n^2$ for large $n$ where $k$ is a constant close to 2. Approximations to $r_{finalpeak}$ can also be obtained by examining its relationship to the positions of the final peaks in truncated functions formed using only last one or more of the terms in the $(R_{n,l}r)^2$ Laguerre-type summation. It is relatively easy to write down explicit expressions for the maxima of these contrived functions. The various maxima in the $(R_{n,l}r)^2$ functions arise as a result of a sensitive balance between successively higher powers of $r$ and their coefficients, the final maximum being a competitive interaction between the ultimate divergence of the Laguerre summation due to its highest order term and the factor of $\exp(-r/(nb_0))$. One might expect therefore to see some empirical relation between $r_{finalpeak}$ of the function $(R_{n,l}r)^2$ and the radial positions of the peaks of the truncated functions $(R_{n,\ell}r)^2_1$, $(R_{n,\ell}r)^2_2$ and $(R_{n,\ell}r)^2_3$, formed using the last one or more terms of the Laguerre summation. Plots of $(R_{n,l}r)^2$, $(R_{n,\ell}r)^2_1$, $(R_{n,\ell}r)^2_2$ and $(R_{n,\ell}r)^2_3$ are shown in Fig. 3. These simplify to be

$$\left(R_{n,\ell}r\right)^2_1 = \frac{4^n \exp\left(-\dfrac{2r}{nb_0}\right)\left(\dfrac{r}{nb_0}\right)^{2n}}{b_0 n^2 (n+l)!(n-l-1)!} \tag{28}$$

$$\left(R_{n,\ell}r\right)^2_2 = \frac{4^n \exp\left(-\dfrac{2r}{nb_0}\right)\left(\dfrac{r}{nb_0}\right)^{2n}}{b_0 n^2 (n+l)!(n-l-1)!}\left(\frac{-2r + nb_0(n-l-1)(n+l)}{2r}\right)^2 \tag{29}$$

$$\left(R_{n,\ell}r\right)^2_3 = \frac{4^n \exp\left(-\dfrac{2r}{nb_0}\right)\left(\dfrac{r}{nb_0}\right)^{2n}}{b_0 n^2 (n+l)!(n-l-1)!} \times$$

$$\left(\frac{8r^2 - 4nb_0\,r(n-l-1)(n+l) + nb_0(n-l-1)(n-l-2)(n+l)(n+l-1)}{8r^2}\right)^2$$

(30)

Formulae for the positions of any maxima of the truncated functions (28) - (30) are exactly determinable from the zeros their differentials. (For example the maximum of (28) occurs at $r = b_0 n^2$ while formulae for the maxima of (29) and (30) have among them solutions that similarly behave as $r = b_0 n^2$ for large $n$ and small $l$.) Fig. 4 shows that the solutions from (28) - (30) for the relevant maxima verify the relationship $r_{finalpeak} \approx k\, b_0\, n^2$ for the particular case $l = 0$. Furthermore, the fact that the solutions approach $r_{finalpeak} \approx k\, b_0\, n^2$ from above and below constrain the value of $k$ to be very close to 2 for large $n$. It seems therefore a reasonable assumption that, when $l = 0$, the high $n$ radial probability density functions $(R_{n,l} r)^2$ extend from near $r = 0$ out to $r_{finalpeak} \approx 2 b_0 n^2$. For any given $n$, as $l$ decreases, the total spread of $(R_{n,l} r)^2$ decreases in a manner approximately proportional to $\sqrt{l}$ until, when $l = n - 1$, the solution given in the previous section applies.

Consideration of $(R_{n,l} r)^2$ for ($l = n$-2), ($l = n$-3), etc., verifies that as $l$ decreases relative to $n$, successive $(R_{n,l} r)^2$ maxima spread out in an approximate square root dependence around this peak. This is further discussed in Appendix B.

# APPENDIX B. APPROXIMATE POSITIONS AND SPACINGS (SPATIAL PERIODS) OF THE LAGUERRE POLYNOMIAL ZEROS FOR ARBITRARY $n$.

Fig. 5 shows the form of a typical radial function $R_{n,l} r$ (from equation (4)) showing its spatially oscillating nature. A little algebra shows that this function, and also the functions $R_{n,l}$ and $(R_{n,l} r)^2$, have the same zeros as the monic polynomial:

$$\Upsilon(g) = \sum_{k=0}^{p-1} \frac{(-1)^{k-p+1} (g)^k (p-1)!(2n-p)!}{(p-1-k)!(2n-2p+k+1)!k!} \qquad (31)$$

where $p = n - l$ and $g = 2r/(nb_0)$ which will be used in the remainder of this appendix. Once the properties of $\Upsilon(g)$ are determined, it is then a straightforward matter to obtain the properties of $R_{n,l}$, $R_{n,l} r$, $(R_{n,l} r)^2$ etc.

Of relevance to the calculation of overlap integrals is the estimation of the spatial extent and position of $\Upsilon(g)$, the spacing (spatial periods) of its zeros, and the height of its peaks. The process begins with an empirical observation shown in Fig. 6 that the set of square roots of the zeros of $\Upsilon(g)$ has a consistent characteristic shape that is approximately linear. The degree of linearity is reasonably independent of the value of $p$ or $n$ but is better when $p \ll n$ and $n \gg 1$ (curve set (d), Fig. 6). $\Upsilon(g)$ is a polynomial of degree $p-1$ and may be written in terms of its $p-1$ zeros, $g_1, g_2, g_3, \ldots g_{p-1}$ as $\Upsilon(g) = (g-g_1)(g-g_2)(g-g_3)\ldots(g-g_{p-1})$. Mathematically, if the square roots of the zeros of $\Upsilon(g)$ are linearly related, then the $i^{\text{th}}$ zero $g_i$ may be written in terms of the first zero $g_1$ as

$$\sqrt{g_i} = \sqrt{g_1} + (i-1)d \qquad (32)$$

where $d$ is a 'spacing' parameter that is constant for a given value of $n$ and $p$. Let the sum of the zeros be $a = \sum_{i=1}^{p-1} g_i$ and the sum of the products of the zeros taken two at a time be $b = \sum_{i,j=1, i \neq j, i > j}^{p-1} g_i g_j = \frac{1}{2}\left(\sum_{j=1}^{p-1}\sum_{i=1}^{p-1} g_i g_j - \sum_{i=1}^{p-1} g_i^2\right)$. Substituting $\sqrt{g_i} = \sqrt{g_1} + (i-1)d$ and $\sqrt{g_i} = \sqrt{g_1} + (i-1)d$ into these summation expressions for $a$ and $b$ and simplifying gives:

$$a = \frac{1}{6}(p-1)\left((2p^2 - 7p + 6)d^2 - 6(p-2)d\sqrt{g_1} + 6g_1\right) \quad (33)$$

$$b = \frac{1}{360}(p-1)(p-2)\begin{pmatrix} (20p^4 - 156p^3 + 427p^2 - 477p + 180)d^4 - \\ 120(p^3 - 6p^2 + 11p - 6)d^3\sqrt{g_1} + \\ 60(5p^2 - 20p + 18)d^2 g_1 - \\ 360(p-2)d\, g_1^{3/2} + 180 g_1^2 \end{pmatrix} \quad (34)$$

From simple polynomial theory $a$ is the coefficient of $g^{p-2}$ and $b$ the coefficient of $g^{p-3}$ in equation (31). Using equation (31) gives

$$a = (2n - p)(p - 1) \quad (35)$$
$$b = (2n - p)(2n - p - 1)(p - 1)(p - 2)/2. \quad (36)$$

Eliminating $a$ and $b$ gives:

$$\frac{1}{6}(p-1)\left((2p^2 - 7p + 6)d^2 - 6(p-2)d\sqrt{g_1} + 6g_1\right) - (2n - p) = 0 \quad (37)$$

$$180(2n - p)(2n - p - 1) = \begin{pmatrix} (20p^4 - 156p^3 + 427p^2 - 477p + 180)d^4 - \\ 120(p^3 - 6p^2 + 11p - 6)d^3\sqrt{g_1} + \\ 60(5p^2 - 20p + 18)d^2 g_1 - \\ 360(p-2)d\, g_1^{3/2} + 180 g_1^2 \end{pmatrix} \quad (38)$$

Rearranging the quadratic equation (37) to make $d$ the subject which on substitution into (38) gives two polynomial equations in $\sqrt{g_1}$. Some lengthy algebra shows these polynomial equations yield the same quartic equation in $g_1$ which simplified gives:

$$\begin{pmatrix} 4(p - 2n)^2 (2p - 3)\left(30 + p(-50 + (37 - 9p)p + n(8p - 19))\right)^2 \\ + 24p(p - 2n)^2 \left(210 + p\begin{pmatrix} -380 + n(191 + p(-303 + (155 - 26p)p)) \\ + p(117 + p(139 + 2p(9p - 50))) \end{pmatrix}\right) g_1 \\ + 4(2n - p)p\left(-90 + p\begin{pmatrix} -(27 + 4p(2p - 7))(-10 + p(9p - 16)) + \\ n(-669 + 2p(596 + p(62p - 335))) \end{pmatrix}\right) g_1^2 \\ + 12p^2(p - 2n)(2p - 1)(11 + p(3p - 11))g_1^3 + (1 - 2p)^2 p^2 (2p - 3)g_1^4 \end{pmatrix} = 0 \quad (39)$$

Although not repeated here because they are rather lengthy, it is immediately possible to solve (39) and hence write down formulae for the approximations to the first ($g_1$) and last ($g_{p-1}$) zeros of, $\Upsilon(g)$ and the spacing parameter $d$, directly in terms of $n$ and $p$. Using (32) then enables approximations of all of the zeros of $\Upsilon(g)$ to be obtained, thus enabling estimation of the spread and extent of $R_{n,l}$ and the separation $\Delta r_i = \left( \mathbf{g}_i - \mathbf{g}_{i-1} \right) n \, b_0 / 2$ of each of its zeros. Approximations are excellent for values of $n$ and $p$ that can be checked by direct numerical solution of the relevant polynomial. For example, with $n = 30$ and $p = 10$, the approximation formulae zero set is {22.0, 27.6, 33.7, 40.5, 47.9, 56.0, 64.6, 73.9, 83.8} while the direct numerical solution zero set of $\Upsilon(\mathbf{g})$ is {22.4, 28.3, 34.3, 40.5, 47.3, 54.8, 63.2, 73.2, 85.9}, that is accuracies of ~ 5% or less. For $n \gg 1$ and $p \ll n$ the accuracy of the approximation is remarkable. Using $n = 1.0 \times 10^{34}$ and $p = 4$, calculations using a precision of 100 significant figures give the radii of the zeros for $R_{n,l}$ at ~ $3.6 \times 10^{21}$ m with all zero spacings of ~ 44 km. Numerical calculation of the zeros of the exact formula for $R_{n,n-4}$ (equation (4)), again using a precision of 100 significant figures, yield zeros that differ from the approximations obtained using the square root spacing assumption by less than 1 part in $10^{18}$ leading to approximations of the zero spacings at these vast distances that differ by less than 4%.

# APPENDIX C. PROCEDURE FOR CALCULATION OF SOME GRAVITATIONAL EIGENSTATE TRANSITION RATES

The matrix element $p_{if}$ can be explicitly written in its components $p_{ifx}$, $p_{ify}$ and $p_{ifz}$ by

$$p_{if} = \sqrt{p_{ifx}^2 + p_{ify}^2 + p_{ifz}^2} \qquad (40)$$

where

$$p_{ifx} = \int_0^\infty \int_0^\pi \int_0^{2\pi} R_{nf,lf}^* Y_{lf,mf}^* \, e \, r \, R_{ni,li} Y_{li,mi} \, r^2 \sin(\theta)\cos(\phi)\sin(\theta) \, d\phi \, d\theta \, dr \qquad (41)$$

$$p_{ify} = \int_0^\infty \int_0^\pi \int_0^{2\pi} R_{nf,lf}^* Y_{lf,mf}^* \, e \, r \, R_{ni,li} Y_{li,mi} \, r^2 \sin(\theta)\sin(\phi)\sin(\theta) \, d\phi \, d\theta \, dr \qquad (42)$$

$$p_{ifz} = \int_0^\infty \int_0^\pi \int_0^{2\pi} R_{nf,lf}^* Y_{lf,mf}^* \, e \, r \, R_{ni,li} Y_{li,mi} \, r^2 \cos(\theta)\sin(\theta) \, d\phi \, d\theta \, dr \qquad (43)$$

The total decay rate for any level will be sum of the transition rates through each of its available decay channels. Finding these transition rates requires calculation of the relevant values of $\omega$ and $p_{if}$, the latter obtained by incorporating the initial and final eigenstate solutions (3) into equations (40) to (43). If $l$ is sufficiently close to $n$ (that is $p\,(\equiv n-l)$ is sufficiently small), then solutions (3) contain a manageable number of terms and it is then possible to obtain decay rates for these specific cases. The relevant states in the galactic halo involve high principal quantum number (large $n$) states.

In Fig. 1 each point represents the $2l+1$ possible *z*-projections of the particular angular momentum state being considered. In this diagram each $p$ value corresponds to a diagonal line starting from the left diagonal $p=1$. The standard dipole selection rules are assumed for transitions between states, $\Delta l = \pm 1$ and $\Delta m = 0, \pm 1$. Thus in Fig. 1, downward transitions originating from states such as A along the first diagonal ($p=1$, $l=n-1$), there is only one decay channel available because of $\Delta l = \pm 1$. For transitions originating from states such as B along the next diagonal ($p=2$, $l=n-2$), two decay channels are possible, with $\Delta n = 1$ or 2. Transitions originating from states like C, E and D further to the right in the figure ($l = n$-3, *n*-4, *n*-5 etc.) have progressively more decay channels available with larger $\Delta n$ values. A significant point to note here however is that each time a transition from an initial state ($l_i$, $n_i$) to a final state ($l_f$, $n_f$) occurs, $n_f - l_f \leq n_i - l_i$. No transitions occur that transfer a particle to a higher $p$ diagonal. The trend therefore is that, as radiative decay proceeds, there is a net migration of states towards the leftmost diagonal at the left of the diagram, that is, to higher $l$ relative to $n$ values. Spontaneous decay favours production of the high $n$, $l$ levels.

Each component of $p_{if}$ may be further split into its separate radial and angular integrals, for example, $p_{ifz} = \int_0^\infty eR^*_{nf,lf}(r)r^3 R_{ni,li}(r)dr \int_0^\pi \int_0^{2\pi} Y^*_{lf,mf} Y_{li,mi} \cos(\theta)\sin(\theta)d\phi d\theta$. A further simplification arises through the explicit calculations of the angular dependency for which it is easily shown that, when $\Delta m = \pm 1$, $p_{ifz}$ is zero and when $\Delta m = 0$, $p_{ifx}$ and $p_{ify}$ are zero.

The angular components, $\int_0^\pi \int_0^{2\pi} Y^*_{lf,mf} Y_{li,mi} \sin(\theta)\cos(\phi)\sin(\theta)d\phi d\theta$,
$\int_0^\pi \int_0^{2\pi} Y^*_{lf,mf} Y_{li,mi} \sin(\theta)\sin(\phi)\sin(\theta)d\phi d\theta$ and $\int_0^\pi \int_0^{2\pi} Y^*_{lf,mf} Y_{li,mi} \cos(\theta)\sin(\theta)d\phi d\theta$
corresponding to $p_{ifx}, p_{ify}$ and $p_{ifz}$, respectively, depend on the initial and final values of $m$. For simplicity only the cases where $l$ and $m$ are even are presented here (similar derivations and results can be undertaken where $l$ and/or $m$ are odd). $Y_{l,m}$ may be written as $Y_{l,m} = \sqrt{(2l+1)(l-m)!/(2(l+m)!)} P_l^m(\cos\theta)$. Introducing the parameter $j = l - m$, using Rodrigues' formula and the binomial theorem, and adjusting the summation limits appropriately, gives the initial state $Y_{li,mi} = Y_{l,m}$ as

$$Y_{l,m} = (-1)^{(l-j)} \sqrt{\frac{2l+1}{4\pi} \frac{j!}{(2l-j)!}} \exp(i(l-j)\phi) \frac{1}{2^l l!}(1-\cos^2\theta)^{(l-j)/2} \times \left\{ \sum_{k=l-j/2}^{l}\left[(-1)^{(l-k)} \frac{(2k)!(\cos\theta)^{(2k-2l+j)}}{(2k-2l+j)!} \frac{l!}{(l-k)!k!}\right] \right\} \tag{44}$$

and the final state $Y_{lf,mf} = Y_{l-1,m}$ with $m = (l-1) - (j-1)$ as

$$= (-1)^{(l-j)} \sqrt{\frac{2l-1}{4\pi} \frac{(l+j)!}{(2l-j-1)!}} \exp(i(l-j)\phi) \frac{1}{2^{(l-1)}(l-1)!}(1-\cos^2\theta)^{(l-j)/2} \times \left\{ \sum_{k=l-j/2}^{(l-1)}\left[(-1)^{(l-k-1)} \frac{(2k)!(\cos\theta)^{(2k-2l+j+1)}}{(2k+j)!} \frac{(l-1)!}{(l-k-1)!k!}\right] \right\} \tag{45}$$

The angular part of $p_{ifz}$ (call it $I_z$) then becomes

$$\int_0^{2\pi} \int_0^{\pi} \begin{pmatrix} (-1)^{(l-j)} \sqrt{\frac{2l+1}{4\pi} \frac{j!}{(2l-j)!}} \exp(i(l-j)\phi) \frac{1}{2^l l!} (1-\cos^2\theta)^{(l-j)/2} \times \\ (-1)^{(l-j)} \sqrt{\frac{2l-1}{4\pi} \frac{(l+j)!}{(2l-j-1)!}} \exp(i(l-j)\phi) \frac{1}{2^{(l-1)}(l-1)!} (1-\cos^2\theta)^{(l-j)/2} \times \\ \left\{ \sum_{k=l-j/2}^{l} \left[ (-1)^{(l-k)} \frac{(2k)!(\cos\theta)^{(2k-2l+j)}}{(2k-2l+j)!} \frac{l!}{(l-k)!k!} \right] \right\} \times \\ \left\{ \sum_{k=l-j/2}^{(l-1)} \left[ (-1)^{(l-k-1)} \frac{(2k)!(\cos\theta)^{(2k-2l+j+1)}}{(2k+j)!} \frac{(l-1)!}{(l-k-1)!k!} \right] \right\} \cos\theta \sin\theta \end{pmatrix} d\theta \, d\phi$$

(46)

Integrating over $\phi$ and putting

$$P_p = \frac{(-1)^p (-1)^{j/2} l! (2p - j + 2l)!}{(p - j/2 + l)!(j/2 - p)!(2p)!} \quad \text{and} \quad Q_q = \frac{(-1)^{q+1} (-1)^{j/2} l! (2p - j + 2l)!}{(q - j/2 + l)!(j/2 - q - 1)!(2q + 1)!} \quad \text{gives,}$$

after adjusting limits, $I_z$ as

$$\frac{1}{2^{2l} l!(l-1)!} \left( \sqrt{\frac{(4l^2 - 1) j!(j-1)!}{(2j-1)!(2l-j-1)!}} \right) \times$$

$$\int_0^{\pi} \cos\theta \sin\theta^{(2(l-j)+1)} \left( \sum_{p=0}^{j/2} P_p \cos^{2p}\theta \right) \left( \sum_{p=0}^{j/2-1} Q_q \cos^{2q+1}\theta \right) d\theta$$

(47)

On expanding the product of the summations and integrating this expression for $I_z$ may be finally reduced to

$$\frac{1}{2^{2l} l!(l-1)!} \left( \sqrt{\frac{(4l^2 - 1) j!(j-1)!}{(2j-1)!(2l-j-1)!}} \right) \times$$

$$\left( \left( \sum_{x=0}^{j/2-1} \left( \frac{2^{2(l-j)} (2x+2)!(l-j)!(2l-2j+1)!(l-j+x+1)!}{(l-j+1/2)(2l-2j)!(x+1)!(2l-2j+2x+3)!} \times \right. \right. \right.$$
$$\left. \left. \left. \left( \sum_{i=0}^{x} \left( \frac{(-1)^{x+1} l!(2i-j+2l)!(l-1)!(2x-2i-j+2l)!}{(i-j/2+l)!(2l-2j)!(2i)!(x-i-j/2+l)!(j/2+i-x-1)!(2x-2i+1)!} \right) \right) \right) + $$
$$\left( \sum_{x=j/2}^{j-1} \left( \frac{2^{2(l-j)} (2x+2)!(l-j)!(2l-2j+1)!(l-j+x+1)!}{(l-j+1/2)(2l-2j)!(x+1)!(2l-2j+2x+3)!} \times \right. \right.$$
$$\left. \left. \left. \left( \sum_{i=x-j/2+1}^{j/2} \left( \frac{(-1)^{x+1} l!(2i-j+2l)!(l-1)!(2x-2i-j+2l)!}{(i-j/2+l)!(2l-2j)!(2i)!(x-i-j/2+l)!(j/2+i-x-1)!(2x-2i+1)!} \right) \right) \right) \right) \right)$$

(48)

The above expression is a monotonically decreasing function of $l$ for all values of $j$ and is $<1$ for all values of $j$ and $l$. In some cases it is easy to write down simple expressions for $I_z$. For example when $j=2$ $I_z = \sqrt{(l-1)/(4l^2-1)}$ so that $I_z \to 0$ as $l \to \infty$. Similar results are obtained when the parameters are odd or combinations of odd and even.

When $\Delta m = \pm 1$ the matrix element involves the components $p_{ifx}$ and $p_{ify}$ because the integral components $I_x = \int_0^\pi \int_0^{2\pi} Y^*_{lf,mf} Y_{li,mi} \sin(\theta)\sin(\phi)\sin(\theta) d\phi d\theta$ and $I_y = \int_0^\pi \int_0^{2\pi} Y^*_{lf,mf} Y_{li,mi} \cos(\phi)\sin(\theta) d\phi d\theta$ are non-zero. Using similar procedures to the above, it can be shown that these integrals are small ($\leq \pi/2$) for all values of $l$ and $j$.

A general form for the radial component of the overlap integral $\int_0^\infty e R^*_{nf,lf}(r) R_{ni,li}(r) r^3 dr$ is given below, although estimates of its behaviour are generally less predictable than those for the angular component and usually require direct computation of each specific case. Furthermore, because of computing limitations (e.g. calculation of $10^{34}$ factorial), the actual numerical calculations were undertaken using a logarithmic approach, and require the use of Stirling's approximation for the factorial function.

The radial component of the initial state is taken as $(n_i, l_i) = (n_i, l) = (n_i, n_i - p)$ where $p = n_i - l$. The final state $(n_f, l_f)$ requires that $l_f = l \pm 1$ taken here as $l_f = l+1$ with a similar result applying for $l_f = l+1$. (Note that the parameters $P_u$ and $Q_v$ used below are not the same as those used in the angular integrals above.) Using the constant $b_0 = \hbar^2/(\mu GmM)$, the initial and final states written in terms of $p$ therefore respectively become, using equation (4)

$$R_{n_i,l} =$$

$$\left(\left(\frac{2}{n_i b_0}\right)^3 \left(\frac{(p-1)!}{2n_i\left((2n_i-p)!\right)^3}\right)\right)^{\frac{1}{2}} \exp\left(-\frac{r}{n_i b_0}\right)\left(\frac{2r}{n_i b_0}\right)^{(n_i-p)} \left(\sum_{k=0}^{(p-1)}(-1)^k \frac{\left((2n_i-p)!\right)^2 \left(\frac{2r}{n_i b_0}\right)^k}{(p-k-1)!(2n_i-2p+k+1)!k!}\right)$$

and

$$R_{n_f,l-1} = \left(\left(\frac{2}{n_i b_0}\right)^3 \left(\frac{(n_f-n_i+p-2)!}{2n_f\left((n_i+n_f-p-1)!\right)^3}\right)\right)^{\frac{1}{2}} \exp\left(-\frac{r}{n_i b_0}\right)\left(\frac{2r}{n_i b_0}\right)^{(n_i-p-1)} \times$$

$$\left(\sum_{k=0}^{p}(-1)^k \frac{\left((n_i+n_f-p-1)!\right)^2 \left(\frac{2r}{n_i b_0}\right)^k}{(n_f-n_i+p-k)!(2n_i-2p+k-1)!k!}\right)$$

(49)

The radial integral $\int_0^\infty e R^*_{nf,lf}(r)\, R_{nili}(r)\, r^3\, dr$ becomes

$$\frac{e}{2}\int_0^\infty \left[\left(\frac{(p-1)!(n_f-n_i+j)!}{n_i n_f \left((2n_i-p)!\right)^3 \left((n_f+n_i-p-1)!\right)^3}\right)^{\frac{1}{2}} n_i^{\left(-\frac{3}{2}+p-n_i\right)} n_f^{\left(-\frac{1}{2}+p-n_i\right)} \exp\left(-\left(\frac{n_f+n_i}{n_i n_f b_0}\right)r\right) \times \left(\frac{2r}{b_0}\right)^{(2n_i-2p+2)} \left(\sum_{u=0}^{(p-1)} P_u \left(\frac{2r}{b_0}\right)^u\right)\left(\sum_{v=0}^{j} Q_v \left(\frac{2r}{b_0}\right)^v\right)\right] dr$$

(50)

where

$$P_u = (-1)^u \frac{\left((2n_i-p)!\right)^2}{n_i^u (p-u-1)!(2n_i-2p+u+1)!u!} \quad \text{and}$$

$$Q_v = (-1)^v \frac{\left((n_i+n_f-p-1)!\right)^2}{n_i^v (p-v+n_f-n_i)!(2n_i-2p+v-+1)!v!}.$$

The product of summations over $u$ and $v$ can be then expanded to give a different summation over new variables $x$ and $s$ which is directly integrable:

$$\frac{e}{2}\int_0^\infty \left[\left[\left(\frac{(p-1)!(n_f-n_i+p)!}{n_i n_f ((2n_i-p)!)^3 ((n_f+n_i-p-1)!)^3}\right)^{\frac{1}{2}} n_i^{\left(-\frac{3}{2}+p-n_i\right)} n_f^{\left(-\frac{1}{2}+p-n_i\right)} \times \right.\right.$$
$$\left.\left[\left\{\sum_{x=0}^{(p-1)}\left(\sum_{s=0}^{x} P_s Q_{x-s}\right)\exp\left(-\left(\frac{n_f+n_i}{n_i n_f b_0}\right)r\right)\left(\frac{2r}{b_0}\right)^{x+2n_i-2p+2}\right\}+\right.\right.$$
$$\left.\left.\left\{\sum_{x=p}^{(2p-1)}\left(\sum_{s=x-p}^{p-1} P_s Q_{x-s}\right)\exp\left(-\left(\frac{n_f+n_i}{n_i n_f b_0}\right)r\right)\left(\frac{2r}{b_0}\right)^{x+2n_i-2p+2}\right\}\right]\right]dr \quad (51)$$

Using the fact that

$$\int_0^\infty \exp\left(-\left(\frac{n_f+n_i}{n_i n_f b_0}\right)r\right)\left(\frac{2r}{b_0}\right)^{x+2n_i-2p+2} dr = \frac{b_0(x+2n_i-2p+2)!}{2}\left(\frac{2n_i n_f}{(n_i+n_f)}\right)^{(x+2n_i-2p+3)}$$

The expression can be finally reduced to

$$\frac{e}{2}\left[\begin{array}{l}\left[\left(\frac{(p-1)!(n_f-n_i+p)!}{n_in_f((2n_i-p)!)^3((n_f+n_i-p-1)!)^3}\right)^{\frac{1}{2}}n_i^{\left(-\frac{3}{2}+p-n_i\right)}n_f^{\left(-\frac{1}{2}+p-n_i\right)}\times\right.\\\left[\left\{\sum_{x=0}^{(p-1)}\left[\left(\sum_{s=0}^{x}\left(\begin{array}{c}(-1)^x\frac{((2n_i-p)!(n_i+n_f-p-1)!)^2}{n_i^x(p-s-1)!(p-(x-s)+n_f-n_i)!}\\\times\frac{1}{(2n_i-2p+(x-s)-1)!(x-s)!(2n_i-2p+s+1)!s!}\end{array}\right)\right)\right.\\\left.\times\frac{b_0(x+2n_i-2p+2)!}{2}\left(\frac{2n_in_f}{(n_i+n_f)}\right)^{(x+2n_i-2p+3)}\right]\right\}+\\\left\{\sum_{x=p}^{(2p-1)}\left[\left(\sum_{s=x-p}^{p-1}\left(\begin{array}{c}(-1)^x\frac{((2n_i-p)!(n_i+n_f-p-1)!)^2}{n_i^x(p-s-1)!(p-(x-s)+n_f-n_i)!}\\\times\frac{1}{(2n_i-2p+(x-s)-1)!(x-s)!(2n_i-2p+s+1)!s!}\end{array}\right)\right)\right.\\\left.\times\frac{b_0(x+2n_i-2p+2)!}{2}\left(\frac{2n_in_f}{(n_i+n_f)}\right)^{(x+2n_i-2p+3)}\right]\right\}\end{array}\right]\right] \quad (52)$$

This expression now gives the radial component of the overlap integral $\int_0^\infty eR_{nf,lf}^*(r)r^3 R_{ni,li}(r)r^2\,dr$, for any general values of $n_i, n_f$ and $p\,(=n_i-l)$. For example substituting $n_f=n_i-1$ and $p=1$ into equation (52) gives an explicit expression for transitions of the type A to A′ ($(n_i=n_i, l=n_i-1)\to(n_f=n_i-1, l=n_i-2)$) shown in Fig. 1 as:

$$\int_0^\infty eR_{nf,lf}^*(r)r^3 R_{ni,li}(r)r^2\,dr = 2^{2n_i} eb_0\left(\frac{n_i(n_i-1)}{(2n_i-1)^2}\right)^{n_i+1}\sqrt{(2n_i-1)^3(2n_i-2)} \quad (53)$$

which, for large $n_i$ becomes $eb_0 n_i^2$.

Eigenstates on the $(l=n-p)^{\text{th}}$ diagonal have radial eigenfunctions that are Laguerre polynomials with $p$ turning points. Consider now transitions of the type $((n_i=n_i, l_i=n_i-2)\to(n_f=n_i-1, l_f=n_i-3))$ taking place along the $p=2$ diagonal of Fig. 1. The initial and final eigenstates involve Laguerre polynomials with 2 turning points. In this case equation (52) yields a summation involving four terms, each of which involves several factorial functions. These may be reduced to the square root of products of terms, which in turn may be simplified using a Taylor expansion to second order around $n_i = 0$. The result gives $eb_0 n_i^2$ for large $n_i$. The radiative decay time is therefore essentially the same as that for transitions of the type A to A′. It can be shown with some difficulty that, whenever transitions take place between two 'p-turning point' Laguerre eigenfunctions, that is, along the same diagonal so that $n \to n-1$ and $l \to l-1$, then the decay rate is $eb_0 n_i^2$ provided $n \gg p$.

Transitions like B to B″ in Fig. 1 involve overlap integrals where the radial component wave functions have very different shapes (B is a 1-turning point function while B″ is a 2-turning point function). As a result, it would be expected in this case that the radial part of the overlap integral would be much smaller than either type A to A′ or B to B′ transitions. Equation (52) gives

$$\int_0^\infty eR^*_{n_f,l_f}(r) r^3 R_{n_i,l_i}(r) r^2 \, dr = 2^{2n_i-2} eb_0 \left( \frac{n_i(n_i-2)}{(2n_i-2)^2} \right)^{n_i} \sqrt{(2n_i-2)(2n_i-3)(2n_i-4)} \quad (54)$$

which reduces to $eb_0 n_i^{3/2} / \sqrt{2}$ again provided $n_i \gg p$.

A general formula for the value of $\int_0^\infty eR^*_{n_f,l_f}(r) r^3 R_{n_i,l_i}(r) r^2 \, dr$ may be obtained for any transition like E to E′ of Fig. 1, which originates from an arbitrary p-turning point radial Laguerre polynomial state $(n_i, l_i = n_i - p)$ and ends on a 1-turning point state $(n_f = n_i - p, l_f = n_f - 1$, lying on the left diagonal). It may be shown that in this case equation (52) reduces to

$$\frac{e}{2} \left[ \left[ \left( \frac{(p-1)!(n_f - n_i + j)!}{n_i n_f ((2n_i - p)!)^3 ((n_f + n_i - p - 1)!)^3} \right)^{\frac{1}{2}} n_i^{\left(-\frac{3}{2}+p-n_i\right)} n_f^{\left(-\frac{1}{2}+p-n_f\right)} \times \right. \right.$$
$$\left. \left. \left\{ \sum_{x=0}^{(p-1)} \left( \left[ \sum_{s=0}^{x} \left( (-1)^x \frac{((2n_i - p)!(n_i + n_f - p - 1)!)^2}{n_i^x (p - s - 1)!(p - (x-s) + n_f - n_i)!} \right. \right. \right. \right. \right. \right.$$
$$\left. \left. \left. \left. \left. \times \frac{1}{(2n_i - 2p + (x-s) - 1)!(x-s)!(2n_i - 2p + s + 1)!s!} \right) \right] \right. \right. \right.$$
$$\left. \left. \left. \times \frac{b_0 (x + 2n_i - 2p + 2)!}{2} \left( \frac{2 n_i n_f}{(n_i + n_f)} \right)^{(x + 2n_i - 2p + 3)} \right) \right\} \right] \quad (55)$$

After some algebraic manipulation, the double summation over **x** and *s* may be simplified and equation (55) becomes

$$\frac{eb_0}{2}\sqrt{(p-1)!}\left(\frac{n_i n_f}{(n_i+n_f)^2}\right)^{(n_f+1)} 2^{2n_i-1}\left(\prod_{i=0}^{p}(n_i+n_f-i)^{\frac{1}{2}}\right)\times$$
$$\left(\sum_{i=1}^{p}\left[(-1)^{i-1}\left(\frac{n_f}{(n_i+n_f)}\right)^i \frac{2^{i-2p+2}(2n_f+i+1)}{(i-1)!(p-i)!}\right]\right) \quad (56)$$

Carrying out the summation over *i* and simplifying the product, (56) becomes

$$eb_0\left(\frac{4n_i n_f}{(n_i+n_f)^2}\right)^{(n_f+1)} \frac{p^{p-2}n_i n_f}{(n_i+n_f)^j}\sqrt{\frac{(n_i+n_f)!}{(p-1)!(2n_f-1)!}} \quad (57)$$

Using Stirling's approximation (57) may be written as

$$eb_0 n_i \left(\frac{4n_i n_f}{(n_i+n_f)^2}\right)^{(n_f+1)}\left(\frac{n_i+n_f+1}{2n_f}\right)^{n_i}\times$$
$$\left(\frac{(n_i+n_f+1)n_f^5}{\mathbf{p}e^2 p^7}\right)^{\frac{1}{4}}\sqrt{\left(\frac{4n_f^2 p}{(n_i+n_f+1)(n_i+n_f)^2}\right)^p} \quad (58)$$

Provided $n_i \gg p$, then $n_i \approx n_f (= n_i - p)$, and using $\lim_{n\to\infty}((2n-p+1)/(2n-2p))^n = \exp((p+1)/2)$, enables (58) to be recast in a form suitable for calculations that involve large values of *n* and *p*:

$$eb_0 n\left(\frac{2}{\mathbf{p}p}\right)^{\frac{1}{4}}\left(\frac{e}{2}\right)^{\frac{p}{2}}\left(\frac{p}{n}\right)^{\frac{p-3}{2}} \quad (59)$$

These important results demonstrate the rapidity with which the value of the radial part of $p_{if}$ decreases as $n_i - n_f$ increases when $n_i$ is large as TABLE I illustrates. The result [*] in TABLE I agrees with the value obtained by the more specialised equation (53).

Once the $p_{if}$ values are known substitution of the other relevant quantities into equation (12) then gives the transition rate for the particular spontaneous transition being considered. From equation (12), the decay rate is proportional to $\mathbf{w}^3 p_{if}^2$. When $n_i$ is large, a transition originating on $p \sim 10^{20}$ might result in an $w^3$ value that could be $10^{75}$ times larger than for a transition from $p = 1$, but $p_{if}^2$ decreases by a much larger factor. The result is that the factor $w^3 p_{if}^2$ rapidly becomes small as $p$ increases, making large $\Delta n$ transitions virtually impossible from the $p = 1$ diagonal. In fact all states on the first million or more diagonals are effectively frozen in time.

The rapid decrease in $p_{if}$ is physically understandable in terms of the shapes and behaviour of the Laguerre polynomials for high $n$, which are obtainable through some of their empirical relationships. When $p$ is large, the polynomial contains many symmetrical oscillations, overlapped with the single peaked and relatively wide $p = 1$, $(l = n - 1)$ polynomial. This oscillatory behaviour of the high $p$ Laguerre functions leads to almost complete cancellation in the overlap integral. Furthermore for large $j$, the width of the polynomial becomes very large and the amplitude consequently very small because of the normalisation condition. For example for $n = 10^{30} + 10^{20}$, $l = 10^{30}$ (i.e. $p = 10^{20}$) the total spread of the wave function is around $3 \times 10^{15}$ m, the wave function amplitude is around $2 \times 10^{-28}$, while the oscillation width is about $3 \times 10^{-5}$ m. The single peaked eigenstate, $n = 10^{30}$, $l = 10^{30} - 1$, (and $p = 1$) has a width of around $6 \times 10^5$ m. There are therefore over ten billion oscillations of the $j = 10^{20}$ state under the single peaked $p = 1$ state.

TABLE I. Radial component integral of $p_{if}$ for transitions of the type $E \rightarrow E'$ of Fig. 1
$(n_i, l_i) \equiv (n, n-p) \rightarrow (n_f, l_f) \equiv (n-p, n-p-1)$

| $n$ | $p$ | $\int_0^\infty eR^*_{nf,lf}(r)r^3 R_{ni,li}(r)r^2\, dr$ (in units of $e\, b_0$) |
|---|---|---|
| 1000 | 1 | $\sim 10^6$ |
| 1000 | 5 | $\sim 6$ |
| 1000 | 20 | $\sim 3 \times 10^{-11}$ |
| 1000 | 100 | $\sim 10^{-40}$ |
| $10^{30}$ | 1 | $\sim 10^{60}$ * |
| $10^{30}$ | 5 | $\sim 6$ |
| $10^{30}$ | 10 | $\sim 10^{-71}$ |
| $10^{30}$ | $10^{20}$ | $\sim 10^{-5 \times 10^{20}}$ |
| $10^{30}$ | $10^{26}$ | $\sim 10^{-2 \times 10^{26}}$ |
| $8 \times 10^{33}$ | $5 \times 10^{31}$ | $\sim 10^{-5 \times 10^{31}}$ |

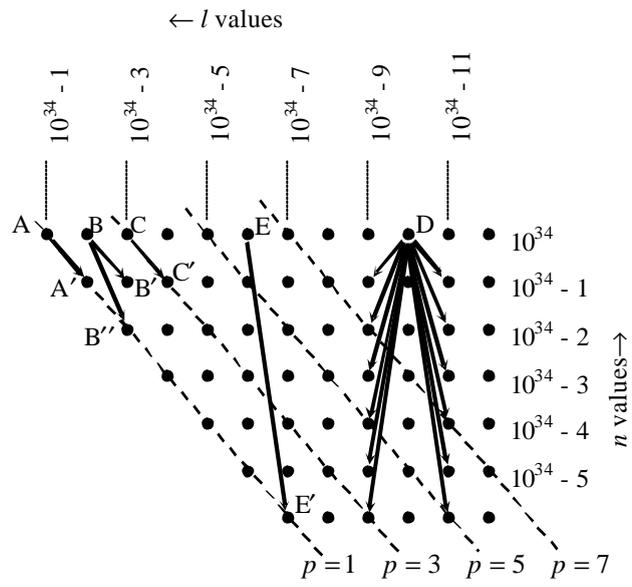

Fig. 1. Points representing the high *n* and *l* valued gravitational macro-eigenstates. Each point represents $(2l+1)$ *z*-projection sublevels.

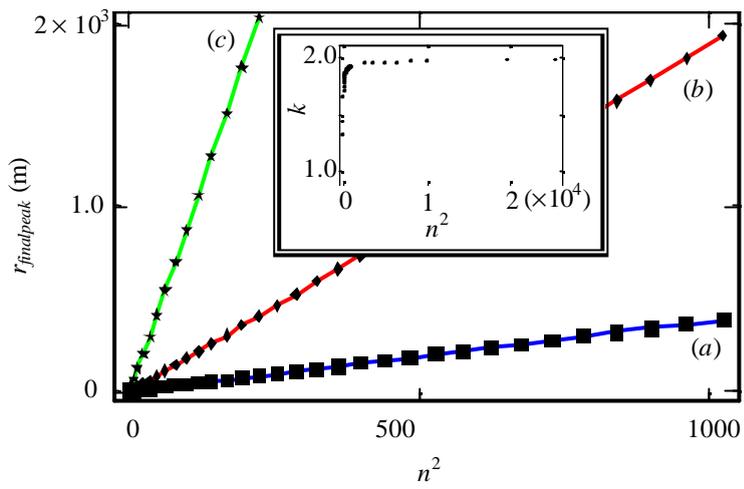

Fig. 2. $r_{finalpeak}$ versus $n^2$ for $n = 1$ to $32$, showing the relationship $r_{finalpeak} \approx k b_0 n^2$. Curve (*a*): $b_0 = 0.2$, curve (*b*) $b_0 = 1.0$ and curve (*c*) $b_0 = 5.0$. The inset shows how the gradient $k$ of curve (*b*) varies with $n^2$.

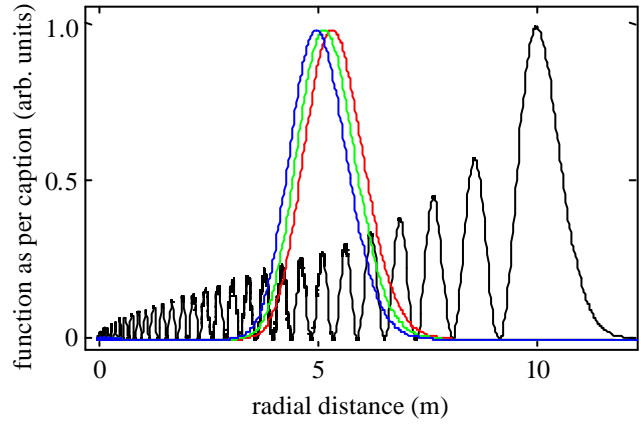

Fig. 3. Plots of the functions $(R_{n,l} r)^2$, $(R_{n,\ell} r)^2_1$, $(R_{n,\ell} r)^2_2$ and $(R_{n,\ell} r)^2_3$ (black, red, green and blue curves, respectively) with normalised peak heights shown as a function of the radial distance $r$ for $n = 32$, $l = 0$, $b_0 = 5.21 \times 10^{-2}$. The peak positions of the truncated functions are approximately half of $r_{finalpeak}$ in agreement with $r_{finalpeak} \approx k b_0 n^2$, $k \approx 2$.

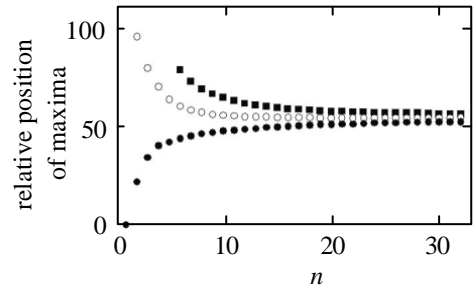

Fig. 4. Relative position of $r_{finalpeak}$ compared to the final peaks of the each of the truncated functions $\left(R_{n,\ell}r\right)^2_1$ (●); $\left(R_{n,\ell}r\right)^2_2$ (■); and $\left(R_{n,\ell}r\right)^2_3$ (○) as a function of $n$ for $l = 0$.

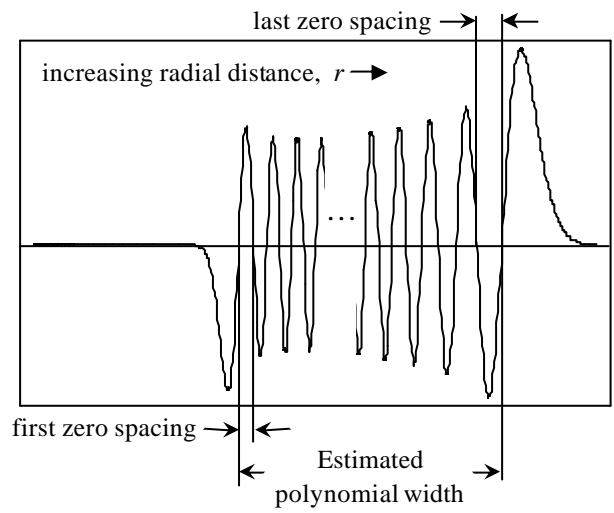

Fig. 5. Schematic diagram showing the typical oscillatory behaviour of the function $R_{nl}(r)\,r$, the first and last zero spacing and the criteria for estimating the Laguerre polynomial width (see Appendix B).

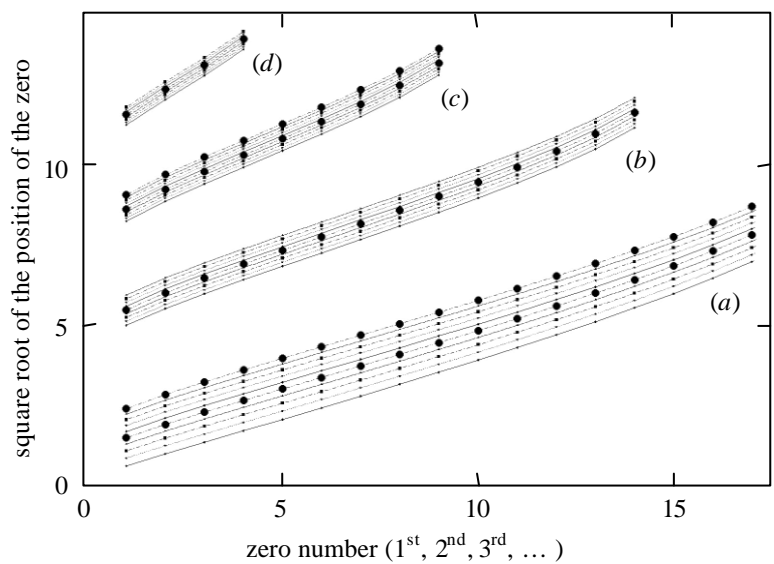

Fig. 6. Square roots of the $1^{st}, 2^{nd}, 3^{rd}$ ... zeros of the function $\Upsilon(g) = \sum_{k=0}^{p-1} \frac{(-1)^{k-p+1}(g)^k (p-1)!(2n-p)!}{(p-1-k)!(2n-2p+k+1)!k!}$

as a function of the zero number.

Curve set (*a*): $p = 20,\ n = \{21,...30\}$; curve set (*b*): $p = 15,\ n = \{41,...50\}$;

curve set (*c*): $p = 10,\ n = \{61,...70\}$; curve set (*d*): $p = 5,\ n = \{81,...89\}$.